\documentclass[12pt,recno]{JHEP3}

\usepackage{graphicx,amsfonts,amssymb,stmaryrd,amsmath,lgdef,longtable}

\title{B-type defects in Landau-Ginzburg models}  
\author{Ilka Brunner$^{1}$ and 
Daniel Roggenkamp$^{2}$ \\  
\\
$^{1}$Institut f{\"u}r Theoretische Physik, ETH Z\"urich \\
CH--8093 Z{\"u}rich, Switzerland\\ 
E-mail: \email{brunner@itp.phys.ethz.ch}\\
\\
$^{2}$Department of Physics and Astronomy, Rutgers University\\
Piscataway, NJ 08855-0849, USA\\
E-mail: \email{roggenka@physics.rutgers.edu}}
\abstract{We consider Landau-Ginzburg models with possibly different
  superpotentials glued together along one-dimensional defect lines. Defects
  preserving B-type supersymmetry can be represented by matrix
  factorisations of the difference of the superpotentials. The
  composition of these defects and their action on B-type boundary
  conditions is described in this framework. The cases of
  Landau-Ginzburg models with superpotential $W=X^d$ and $W=X^d+Z^2$
  are analysed in detail, and the results are compared to the CFT
  treatment of defects in  $N=2$ superconformal minimal models to
  which these Landau-Ginzburg models flow in the IR.}

\preprint{}

\begin{document}
\section{Introduction}
Over the last decades, D-branes and their descriptions from
various points of views such as boundary states in conformal field theory
or solutions to supergravity equations have received a lot of attention. 
An interesting and quite useful property of D-branes is that there exist
operations which act naturally on them and relate D-branes in
possibly different theories. Examples of such operations are dualities
such as T-duality or mirror symmetry, but also 
monodromy transformations along paths in moduli spaces. Some of those
operations, namely the natural operations on the category of
B-type D-branes in Calabi-Yau compactifications, have
an elegant description in terms of Fourier-Mukai transformations.
All these examples have in common that they do not involve the string
coupling $g_s$ and hence can be studied at weak coupling
in the framework of conformal field theory.

From a world sheet point of view, a natural operation on D-branes
is provided by defects, one-dimensional interfaces, along which 
two possibly different conformal field theories are glued together. 
Defects in conformal field theory have received
some attention recently 
\cite{Petkova:2000ip,Bachas:2001vj,Graham:2003nc,Frohlich:2004ef,Bachas:2004sy,Frohlich:2006ch,Quella:2006de,Fuchs:2007tx,Fuchs:2007fw}
and have also emerged as domain walls in the discussion of the 
${\rm AdS}_3/{\rm CFT}_2$-duality \cite{Bachas:2001vj}.

Among the defects in conformal field theories, there is a special
class of so-called {\it topological} defects, which have the property
that they can be moved freely across the world sheet, as long as they do not
cross field insertions or other defects. They act naturally on
conformal boundary conditions, \ie D-branes. Namely, when both
defects and boundaries are present one can bring 
defects close to the world sheet boundary and in the limit in which
the defect approaches the boundary, a new boundary
condition arises. To put it differently, the world sheet boundary couples to
a new D-brane. 
Likewise, one can bring together two topological defects. In the
limit in which the two merge, one obtains  new defects, and hence
topological defects can be composed \cite{Petkova:2000ip,Frohlich:2006ch}.

Generic defects however cannot be composed
or act on boundary conditions in this way. 
In general, correlation functions in the
presence of defects depend on the positions of the latter, and
in particular exhibit singularities when defects approach each other or
world sheet boundaries. So the process of merging defects, or moving
defects to world sheet boundaries is a priori not well defined for generic
conformal defects. 

Of course, defects can also be studied in the context of topological
rather than conformal field theory. 
In topological field theories, where correlation functions do not depend on the
world sheet metric, defects can always be moved  and
hence can always merge and act on the D-branes of the 
topological theory. Therefore,
{\it any} defect in topological field theory provides a suitable map
between the respective D-brane categories. 

This is in particular true for defects in topologically twisted
$N=(2,2)$ supersymmetric field theories (including superconformal
field theories). As will be described in Section \ref{sectgeneral}, 
supersymmetry preserving defects in $N=(2,2)$ supersymmetric field
theories come in two variants, just as D-branes or orientifolds do. 
A-type defects are compatible with the topological A-twist, B-type
defects with the topological B-twist. Hence the topological twist
endows both of these types of defects with a composition and action on
the respective class of D-branes, even though they are not topological
in the untwisted field theory. Unlike D-branes or orientifolds,
defects can be of A- and B-type at the same time. Those defects are
topological already in the untwisted theory.

Our main focus in this paper will be the investigation of B-type defects
in $N=(2,2)$ supersymmetric Landau-Ginzburg models. These flow to
superconformal field theories in the IR, and play an important role in
the study of string compactifications on Calabi-Yau manifolds, where
they provide useful descriptions of the small volume regime.

We consider the situation where a Landau-Ginzburg model with
superpotential $W_1$ is separated by a defect
from a Landau-Ginzburg model with superpotential $W_2$. We argue 
that similar to B-branes in these models, which can be represented by
matrix factorisations of the $W_i$ \cite{Kontsevich,Kapustin:2002bi,Brunner:2003dc}, 
B-type defects between the models are described by matrix factorisations of the 
difference $W_1-W_2$ of the superpotentials 
(see \cite{Khovanov:2004bc,Kov-Roz} for earlier work that has discussed
these defects from a slightly different point of view). 
We then give a prescription of the composition of these defects, and
their action on the respective boundary conditions in this framework.

We will discuss in particular a simple class of defects, which do not
introduce any additional degrees of freedom. Such defects are related
to symmetries and indeed orbifolds of the underlying bulk theories.
They implement the action of these symmetries on bulk
fields, and their defect-changing operators correspond to twisted sectors in
the respective orbifold models. 

For the simple classes of Landau-Ginzburg models with a single chiral
superfield and superpotential $W=X^d$ and their 
cousins\footnote{Although the bulk Landau-Ginzburg 
  theory hardly changes when adding a square to the superpotential,
  the B-brane spectra of the two theories are rather different
\cite{Kapustin:2003rc}.} with an additional superfield and
superpotential $W=X^d+Z^2$, we compare the description of 
B-type defects, their composition and action on B-type boundary
conditions in the framework of matrix factorisations with the
respective conformal field theory description available in the IR. 
Namely, these Landau-Ginzburg models
flow to $N=2$ superconformal minimal models in the IR, in which
defects can be studied by means of CFT techniques. We find complete
agreement between the two approaches.

This paper is organised as follows. In  Section \ref{sectgeneral}
we set the stage and discuss general properties of 
supersymmetry preserving defects in theories with $N=(2,2)$
supersymmetry. 
Section \ref{sectlgaction} is devoted to the study
of defects in Landau-Ginzburg models.
In particular, we show that supersymmetry preserving B-type defects in 
Landau-Ginzburg models are
described by matrix factorisations. As a next step, in Section
\ref{secoperation} we consider situations
where several defects or both defects and boundaries are present, and  
work out the composition of defects and their action on boundary
conditions in this framework. The special class of symmetry defects is discussed in
Section \ref{sectgrouplike}. Section \ref{comparison} contains the
explicit comparison of B-type defects 
in Landau-Ginzburg with superpotentials $W=X^d$, $W=X^d+Z^2$ 
and defects in the corresponding superconformal 
minimal models. Some technical details appear in the Appendix.
\section{Defects in $N=2$ theories}\label{sectgeneral}
In this paper we consider two-dimensional field 
theories with $N=2$ supersymmetry for both left
and right moving degrees of freedom. There are hence four anti-commuting
supercharges $Q_\pm$, $\bQ_\pm$ satisfying the usual anti-commutation relations
\beq
\{ Q_\pm, \bQ_\pm \} = H \pm P\,,
\eeq
with all other anti-commutators vanishing.
$H$ and $P$ denote energy and momentum density, the superscripts
${}^\pm$ distinguish left and right movers and a bar indicates
conjugation. 

We are interested in situations where two such theories
are glued together along a common one-dimensional interface, a
{\it defect}. 
Our focus will be on supersymmetry preserving defects,
{\it i.e.} those defects whose presence still allows the total theory
to be supersymmetric with respect to one half of the supersymmetries
present in the original bulk theories. 
Just like in
the case of $N=2$ theories on surfaces with boundaries or unoriented
surfaces, there are two ways of doing so. The respective defects are
called {\it A-} and {\it B-type} respectively. Modelling the defect on the real line
$\RR\subset\CC$ separating two possibly different theories on the upper
and lower half plane, {\it B-type defects} have the property that the
combination 
$Q_B=Q_+ + Q_-$ of supercharges and its conjugate $\bar Q_B$ 
are preserved everywhere on $\CC$. 
That means that along the interface $\RR$ the supercharges have to satisfy the
following ``gluing conditions'':
\beqa\label{Bdef}
Q_+^{(1)} + Q_-^{(1)} &=&  Q_+^{(2)}+ Q_-^{(2)}, \\ \nonumber
\bQ_+^{(1)}+ \bQ_-^{(1)}  &=& \bQ_+^{(2)}+ \bQ_-^{(2)}\ .  
\eeqa
Here, the superscripts ${}^{(1)}$ and ${}^{(2)}$ refer to the two theories on upper
and lower half plane respectively.
For {\it A-type defects} on the other hand, the gluing conditions
along the defect are
twisted by the automorphism of the supersymmetry algebra which
exchanges $Q_\pm$ with $\bar Q_\pm$: 
\beqa\label{Adef}
Q_{+}^{(1)} + \bQ_-^{(1)} &=& Q_{+}^{(2)} + \bQ_-^{(2)} , \\ \nonumber
\bQ_{+}^{(1)} + Q_-^{(1)} &=& \bQ_{+}^{(2)} + Q_-^{(2)} .
\eeqa
They ensure that 
the combination $Q_A= Q_+ + \bQ_-$ and its conjugate $\bar Q_A$ are
preserved\footnote{Of course, there are also other automorphisms of the
$N=(2,2)$ supersymmetry algebra, which can be used to twist the gluing
conditions. For $\eta^\pm\in\{\pm 1\}$, $Q_\pm\mapsto\eta^\pm Q_\pm$,
$\bar Q_\pm\mapsto\eta^\pm \bar Q_\pm$ gives rise to modified A- and
B-type gluing conditions. For simplicity of presentation we will
refrain from spelling out the details of these additional
possibilities here, but we will comment on $\eta^\pm$-twisted
gluing conditions in the context of conformal
field theory in Section \ref{CFTapproach}.}.

In situations where defects as well as boundaries 
are present, A- or B-type supersymmetry can be preserved in case all defects and all boundaries
are of A- or B-type respectively.
Just as for D-branes,  mirror symmetry exchanges A- and B-type defects.

Note that there are two special classes of defects which actually preserve the
full $N=(2,2)$ algebra. The first class consists of defects such that
\beq
Q_{\pm}^{(1)} = Q_{\pm}^{(2)}, \quad \bQ_{\pm}^{(1)} = \bQ_{\pm}^{(2)} \quad
{\rm on} \ \RR \ ,
\eeq
which implies both A- as well as B-type gluing conditions \eq{Adef}, \eq{Bdef}.
One particular defect of this kind is the trivial defect between one
and the same theory. 
Defects of the second class are related to those of the first
class by mirror symmetry. They obey the respective mirror twisted gluing conditions
\beqa
Q_{+}^{(1)} &=&  Q_{+}^{(2)}, \quad \bQ_{+}^{(1)} = \bQ_{+}^{(2)} \\ \nonumber
Q_{-}^{(1)} &=& \bQ_{-}^{(2)}, \quad \bQ_{-}^{(1)} = Q_{-}^{(2)} \quad
{\rm on} \ \RR \ .
\eeqa
Such defects exist for example between a theory and its mirror and hence
realise mirror symmetry as a defect.

Using the supersymmetry algebra, it follows immediately that defects
of these two classes 
preserve translational invariance in space and time because the gluing
conditions for the supercharges imply
\beq\label{Hpreserve}
H^{(1)} = H^{(2)}, \quad P^{(1)} = P^{(2)}\quad {\rm on}\;\RR \ .
\eeq
This is not possible for world sheet boundaries which automatically
break one half of the local translation symmetries and therefore can
at most preserve half of the bulk supersymmetries. In contrast, 
a theory with a
defect allows for the possibility of being invariant
under shifts of the defect on the world sheet.

Nevertheless, the similarities between defects and boundaries are
indeed very useful for the treatment of defects. In particular, one can 
obtain an equivalent description of the situation described above
by folding the world sheet along the real line and 
realising the degrees of freedom of the theories on the upper and
lower half plane as different sectors in a ``doubled'' theory defined on the
upper half plane only \cite{Bachas:2001vj,Oshikawa:1996dj}. 
Folding the theory from the lower to the upper
half plane, left and right movers are interchanged, and defects
in the original theory on the complex plane become boundary conditions
in the doubled theory.
If the defect preserves the full $N=(2,2)$ 
supersymmetry, the corresponding boundary conditions in the doubled theory are
of permutation type, \ie left movers of the supercharges in one sector
are glued to right movers of the respective supercharges in the other
one and vice versa. 

Of particular interest in the context of string theory are theories with
$N=(2,2)$ superconformal symmetry. The corresponding symmetry 
algebra is generated by the modes
of the energy momentum tensor $T$, ${\rm U}(1)$-current $J$ and two
supercurrents $G^\pm$ together with the ones of the respective right movers 
$\overline T$, $\overline J$, ${\overline G}^\pm$.
(As is customary in CFT, the superscripts ${}^\pm$ specify the ${\rm
  U}(1)$-charge of the respective current, and right movers will be
distinguished from left movers by a bar. This differs from the
notation used for the supercharges $Q$ in the discussion above.)

In these theories, one can consider defects preserving one half of the bulk
superconformal symmetry. As before, we call them A- and B-type
depending on which combinations of supercharges are conserved. The
corresponding gluing conditions along the real line are given by
\beqa
T^{(1)}-\overline{T}^{(1)} &=& T^{(2)}-\overline{T}^{(2)}\,,\\
J^{(1)} - \overline{J}^{(1)}&=& J^{(2)} - \overline{J}^{(2)}\,,\nonumber\\
G^{\pm(1)} + \overline{G}^{\pm(1)} &=& G^{\pm(2)} +
\overline{G}^{\pm(2)}\nonumber
\eeqa
for B-type defects
and 
\beqa
T^{(1)}-\overline{T}^{(1)} &=& T^{(2)}-\overline{T}^{(2)}\,,\\
J^{(1)} + \overline{J}^{(1)}&=& J^{(2)} + \overline{J}^{(2)}\,,\nonumber\\
G^{\pm(1)} + \overline{G}^{\mp(1)} &=& G^{\pm(2)} +
\overline{G}^{\mp(2)}\nonumber
\eeqa
for A-type defects.

Just as in the general situation, there is also a class of defects that
preserves the full $N=(2,2)$ superconformal symmetry, and is hence both of A- as well as
B-type. Because of the two automorphisms of the $N=2$ superconformal algebra
there are essentially four possible gluing conditions for those
defects. Namely, for $a,\bar a\in\{\pm 1\}$ we have
\beqn
T^{(1)}-T^{(2)}=&0&=\overline{T}^{(1)}-\overline{T}^{(2)}\,,\\
J^{(1)} - a J^{(2)}=&0&=\overline{J}^{(1)}-\bar a
\overline{J}^{(2)}\,,\nonumber\\
G^{\pm(1)} + G^{a\pm(2)}=&0&=
\overline{G}^{\pm(1)}+\overline{G}^{\bar a\pm(2)}\,.\nonumber
\eeqn
These defects in particular glue together holomorphic and
antiholomorphic energy momentum tensors separately and therefore
preserve both holomorphic and antiholomorphic Virasoro algebras. This implies that despite the
presence of such a defect, correlation
functions are still covariant with respect
to all local conformal transformations of the world sheet, 
even those which change the position of the defect. This implies 
that correlation functions do not change when such defects
are shifted on the world sheet. Defects which have this property have
been called {\it topological} in
\cite{Bachas:2004sy}.

Since they can be shifted on the world sheet, topological defects can
in particular be brought on top of each other, to ``fuse'' to new
defects. This procedure furnishes the topological defects with a
composition. Moreover, they can also be brought on top of world
sheet boundaries producing new boundary conditions in this way. Hence,
topological defects act on boundary conditions. This is not true for
non-topological defects. Letting two of those defects approach each
other, or one of them approach a boundary will in general lead to
singularities in correlation functions.

Certain $N=(2,2)$ supersymmetric field theories, in particular those
considered here can be topologically twisted
\cite{Witten:1991zz}. Twisting
changes the energy momentum tensor of the theory in such a way that it
is exact with respect to a BRST-operator $Q=Q_A$ for an A-twisted
theory or $Q=Q_B$ for a B-twisted theory. A consequence of this is
that correlation functions only involving $Q$-closed fields are
invariant with respect to variations of the world sheet metric and
thus define a topological field theory. This twisting procedure is
compatible with the existence of boundaries and defects, as long as
the chosen BRST-charge is preserved by the boundaries and
defects. More precisely, A- and B-twisting is compatible with A- and B-type
boundary conditions and defects respectively. 
Indeed, by arguments similar to those
used for pure bulk theories it follows
that also correlation function of BRST-closed fields in the presence
of boundaries and defects become topological in the twisted theory. 
In particular, upon twisting all defects, even those which have not
been topological in the original untwisted theory become topological,
\ie in the topologically twisted theory they can be shifted on the
world sheet. 
Thus, the topological twisting provides a composition of all A- and
B-type defects and an action of them on A- and B-type boundary
conditions respectively. We will study the action of B-type defects in
Landau-Ginzburg models and their action on B-type boundary conditions
below.

Let us close this section with a few general remarks about
defects. Similarly to boundary conditions, defects add to the
structure of the underlying bulk theories. For instance, if a defect
is located on a closed curve, such that the world sheet can be cut open on
both sides of it, the defect provides a homomorphism between the 
bulk Hilbert spaces of the theories it separates. 
(This is similar to boundary
  conditions giving rise to boundary states.) These defect operators
  are often a convenient way to encode part of the information about a
  defect, and we will make use of it below. 
Note however that there is more structure. Similar to  boundary
conditions which come with additional degrees of freedom such as 
boundary condition changing
boundary fields (open strings between the respective D-branes),  
also defects introduce new degrees
of freedom. Unlike boundaries however, defects can form junctions, and
there are fields localised on all possible junctions of defects
(including the one-junction, which is a tip of a defect). In a more
string theoretic language one would call these degrees of freedom
closed strings twisted by the respective defects. If moreover there are
boundaries and defects in a theory, defects can also end on
boundaries, giving rise to even more degrees of freedom etc. 
Part of these structures will be described in explicit examples
below. 
\section{Defects in Landau-Ginzburg models}\label{sectlgaction}
\subsection{Bulk action}

Our conventions
for the $N=(2,2)$ superspace are those of \cite{Hori:2000ic}. The two-dimensional
$(2,2)$ superspace is spanned by two bosonic coordinates $x^\pm=x^0\pm x^1$ and 
four fermionic coordinates $\theta^\pm$, $\bt^\pm$.
The supercharges are realised as the following differential
operators on superspace
\beq \nn
Q_\pm = \frac{\partial}{\partial \theta^\pm} + i \bt^\pm
\partial_\pm \ ,\qquad
\bQ_\pm = -\frac{\partial}{\partial \bt^\pm} - i \theta^\pm \partial_\pm\,.
\eeq
The superderivatives are given by
\beq\nn
D_\pm = \frac{\partial}{\partial \theta^\pm} - i \bt^\pm \partial_\pm \ ,\qquad
\bD_\pm = -\frac{\partial}{\partial \bt^\pm} + i \theta^\pm \partial_\pm\,.
\eeq
Chiral superfields $X$ satisfy the conditions
$\bD_\pm X =0$ and  have an expansion
\beq
X= \phi(y^\pm) + \theta^\alpha \psi_\alpha(y^\pm) + \theta^+ \theta^-
F(y^\pm)
\eeq 
into components, where $y^\pm= x^\pm - i \theta^\pm \bt^\pm$ and $\alpha \in\{\pm\}$.
The conjugate fields $\bar X$  are anti-chiral, \ie they satisfy $D_\pm
\bar X = 0$. 

We consider Landau-Ginzburg models with a finite number of chiral
superfields $X_i$ and action given by the sum
\beq\label{lgaction}
S=S_D+S_F 
\eeq
of D- and F-term. The D-term
\beq
S_D=\int d^4\theta d^2 x K(X_i,\bar{X}_i) 
\eeq
is determined by the K\"ahler potential $K$ which we will assume to be
flat and diagonal, $K= \sum_i\bar X_i X_i$. 
In the topologically twisted theory, the variation of a D-term is BRST trivial
and therefore all correlation functions are independent of D-term changes.
This is well-known for the case of world sheets without boundary, and has been
extended to world sheets with boundary in \cite{Lazaroiu:2003zi} .
The F-term
\beq
S_F=\int d^2x 
d\theta^- d\theta^+ W(X_i) 
\vert_{\bt^\pm=0} \  + \int d^2 x d\bt^+ d\bt^- \overline{W}(\bar X_i) 
\vert_{\theta^\pm=0} 
\eeq
is parametrised by the superpotential $W$, a holomorphic function of
the chiral superfields $X_i$. 
It is this term, which completely determines the B-twisted model, and we will therefore
focus our discussions on it.
In the case that $W$ is quasi-homogeneous the Landau-Ginzburg model will flow
to a conformal field theory in the IR. 
According to standard arguments, the D-term will adjust itself in this
process to be compatible with the conformal symmetry, whereas the
F-term remains unrenormalised.
Therefore, for  comparisons with conformal field theory 
only the F-term  will be 
relevant.

On a world sheet without boundaries or defects, the Landau-Ginzburg action is manifestly
$N=(2,2)$ supersymmetric, \ie the variation of the action with
respect to 
\beq\label{delta}
\delta = \epsilon_+ Q_- - \epsilon_- Q_+ - \bar\epsilon_+ \bQ_-+ \bar\epsilon_-
\bQ_+ 
\eeq
vanishes for all $\epsilon_\pm$, $\bar\epsilon_\pm$. The corresponding
conserved supercharges can be realised as
\beqn\label{lgsupercharges}
Q_\pm &=& \int dx^1 \big( (\partial_0\pm \partial_1 \big) \bar\phi_{\bar{j}} \psi_\pm^j
\mp i \bar\psi_\mp^{\bar{i}} \partial_{\bar{i}} \overline{W} \big)\,, \\ 
\bQ_\pm &=& \int dx^1 
\big( \bar\psi^{\bar{j}}_\pm(\partial_0\pm \partial_1 \big) \phi_{j} 
\pm i \psi_\mp^{i} \partial_{i} W \big)\,. \nn
\eeqn
\subsection{B-type boundary conditions and matrix factorisations}\label{sectmfbd}
Let us briefly review the formulation of a Landau-Ginzburg theory on the upper half
plane (UHP) 
\cite{Warner:1995ay,Kontsevich,Orlov,Kapustin:2002bi,Brunner:2003dc}.
We will consider the situation where superspace acquires a B-type
superboundary with coordinates
\beq
x^+=x^-=t, \quad \theta^+ = \theta^- = \theta, \quad \bt^+ = \bt^- =\bt \ .
\eeq
The presence of the boundary reduces the number of supersymmetries of
the theory, because only the combinations
\beq\label{deltaB}
\delta_B= \epsilon Q - \bar\epsilon \bQ,
\eeq
of the supersymmetry generators with 
\beq
Q= Q_+ + Q_- , \qquad \bQ = \bQ_+ + \bQ_-
\eeq
are compatible with the B-type boundary. To put it differently, a
supersymmetry of the form \eq{delta} only preserves the 
boundary if $\epsilon_+= -\epsilon_-=:\epsilon$ and
$\bar\epsilon_+ =-\bepsilon_-=:\bar\epsilon$. 

As it turns out, the restriction of the bulk Landau-Ginzburg action to
a world sheet with B-type boundary on its own is not invariant under
the B-type supersymmetry \eq{deltaB}. Namely, 
the $\delta_B$-variation of the bulk Landau-Ginzburg action
\eq{lgaction} in the presence of the boundary
introduces boundary terms
\beq\label{bulkvar}
\delta_B S = \delta_B S_D +\delta_B S_F\,,
\eeq
where in particular the variation of the F-term yields \cite{Warner:1995ay,Hori:2000ic}
\beq\label{bulkvarF}
\delta_BS_F =i \int_{\partial \Sigma} dt d\theta \bepsilon W
-i \int_{\partial \Sigma} dt d\bt \epsilon \bW\,.
\eeq
Thus, in order to define a supersymmetric theory on a surface with boundary, one
either has to impose boundary conditions on the fields, 
which ensure the vanishing of \eq{bulkvar}, or  add an additional
boundary term to the action whose supersymmetry variation compensates
for the term coming from the bulk variation. 
In fact, it has been argued in \cite{Kapustin:2002bi,Brunner:2003dc} that the
D-term in \eq{bulkvar} can always be compensated by the supersymmetry
variation of an appropriately
chosen boundary term, and that the F-term \eq{bulkvarF} can be cancelled by introducing extra
non-chiral fermionic boundary superfields $\pi_1, \dots, \pi_r$
satisfying
\beq
\bD\pi_i = E_i \ .
\eeq
Indeed, the supersymmetry variation of the boundary F-term
\beq\label{bdaction}
\Delta S = i\int_{\partial \Sigma} dt d\theta J_i \pi_i \ + \  c.c
\eeq
exactly cancels the term \eq{bulkvarF}
resulting from the supersymmetry variation of the bulk F-term if 
\beq\label{factorize}
\sum_i J_i E_i = W\,.
\eeq
Therefore, any factorisation \eq{factorize} of the
superpotential $W$ gives rise to a supersymmetric action of the
Landau-Ginzburg model on a surface with B-type boundary.
To put it differently, such a factorisation 
defines a supersymmetric B-type
boundary condition of the model. We will omit the discussion of the kinetic
terms for the boundary fermions, since they play no role in the current
context.

Physically, the D-branes constructed in this way are composites
of a brane-anti-brane pair obtained by a tachyon condensation. To be more 
precise, the brane-anti-brane pair is a pair of flat space-filling
D-branes in the theory with vanishing superpotential $W=0$
--- a sigma-model with target
space $\CC^N$, where $N$ is the number of chiral superfields. The tachyon
condensation is triggered by turning on the superpotential $W$.
In this picture, fermionic degrees of freedom correspond to strings
stretching from brane to anti-brane. In particular, the fermionic matrix $Q$
contains the tachyon profile on the space-time filling brane-anti-brane pair.

As in the case of Landau-Ginzburg theories on surfaces without boundary, one can
also perform a topological twist in the presence of
supersymmetric boundaries to extract information about the
topological sectors. However, only the topological B-twist is
compatible with B-type boundary conditions. The BRST-charge of the
corresponding B-model in the presence of the B-type boundary condition
defined by \eq{factorize} then receives a boundary contribution 
\beq\label{BRSTop}
Q_{\rm bd}= \sum_i J_i \pi_i + E_i \bpi_i \ ,
\eeq
which obeys $Q_{\rm bd}^2=W$ by means of the factorisation condition
(\ref{factorize}). As usual, the degrees of freedom of the twisted
theory are given by the cohomology of the BRST-operator. In
particular, the topological boundary fields are described
by the cohomology of the boundary BRST-operator, which acts on
boundary fields by the graded commutator with $Q_{\rm bd}$. (The
$\ZZ_2$-grading is due to the presence of bosonic and fermionic
degrees of freedom on the boundary.)

Two such $B$-type boundary conditions defined by boundary
BRST-charges $Q_{\rm bd}$ and $Q\p_{\rm bd}$, are equivalent, if there
are homomorphisms $U$ and $V$ between the respective spaces of boundary fields, which preserve
the $\ZZ_2$-grading such that 
\beq\label{Qequiv}
Q\p_{\rm bd}=UQ_{\rm bd}V\,,\qquad UV=\id\p+\{Q\p_{\rm
bd},O\p\}\,,\quad VU=\id+\{Q_{\rm bd},O\}
\eeq
for some $O$ and $O\p$. 
$U$ corresponds to an open string operator propagating from
one brane to the other which can be composed with the ``inverse'' $V$ 
propagating in the other direction to yield the identity operators on
both of the individual branes.

Note that the notion of equivalence in the B-brane category only
requires $U$ and $V$ to be inverse up to BRST-trivial terms\footnote{
It is also possible to consider related categories, in which
the notion of equivalence is different from the one used here. 
For instance instead of taking
as morphism spaces the BRST-cohomology, one could use the space
of BRST-closed operators. In order to define an equivalence in this
category $U$ and $V$ would have to be
genuine inverses of each other.}.
One
consequence of this is that all physically trivial matrix factorisations, \ie
those associated to D-branes which do not have any non-trivial open strings ending on them,
are mutually equivalent. One particular representative of this trivial
factorisation can be obtained by setting $r=1$, $J_1=1$ and
$E_1=W$.
In the language of the covering theory with $W=0$, this amounts to a trivial
brane-anti-brane pair.
Adding it to any other boundary condition
does not change the physical content, and hence gives rise to an
equivalent boundary condition:
\beq
Q_{\rm bd} \sim Q_{\rm bd} \oplus Q_{triv} \ .
\eeq

Choosing an explicit matrix representation of the Clifford algebra
generated by the boundary fermions $\pi_i$, the boundary BRST-charges 
$Q_{\rm bd}$ are represented by
$2^{r+1}\times 2^{r+1}$-matrices of the form
\beq
Q_{\rm bd}=\left(\begin{array}{cc} 0 & p_1 \\ p_0 &
    0\end{array}\right)\,,
\eeq
where the $p_i$ are $2^r\times 2^r$-matrices whose entries are
polynomials in the chiral fields $X_i$ such that
$p_1p_0=W(X_i)\id_{2^r\times 2^r}=p_0p_1$. The $p_i$ constitute a {\it matrix
  factorisation} of $W$ of rank $2^{r}$ and determine $Q_{\rm bd}$ and
hence the B-type boundary condition.  More generally also $Q_{\rm
  bd}$ constructed out of matrix factorisations $p_1,p_0$ of arbitrary 
rank $N$ define meaningful boundary conditions. This has been shown in 
\cite{Lazaroiu:2003zi} by taking into account the gauge degrees of
freedom in higher multiplicity brane configurations in the underlying
$\CC^N$-sigma model.

One often represents matrix factorisations in the following way \cite{Orlov}
\beq\label{matrixfactorisations}
P:\quad P_1=\CC[X_i]^N\overset{p_1}{\underset{p_0}{\rightleftarrows}} \CC[X_i]^N=P_0\,,\qquad
p_1p_0=W(X_i)\id_{P_0}\,,\quad 
p_0p_1=W(X_i)\id_{P_1}\,.
\eeq
As follows from \eq{Qequiv}, two such matrix factorisations $P$ and $P\p$
lead to equivalent boundary conditions, if
there exist homomorphisms $u_i:P_i\rightarrow P\p_i$, $v_i:P\p_i\rightarrow
P_i$ such that
\beq\label{Mequiv}
p\p_1=u_0p_1v_1\,,\quad p\p_0=u_1p_0v_0\,,\quad
p_1=v_0p\p_1u_1\,,\quad p_0=v_1p\p_0u_0
\eeq
and
\beqn
&&v_0u_0=\id_{P_0}+\chi_1p_0+p_1\chi_0\,,\quad
v_1u_1=\id_{P_1}+p_0\chi_1+\chi_0p_1\,,\\
&&u_0v_0=\id_{P\p_0}+\chi\p_1p\p_0+p\p_1\chi\p_0\,,\quad
u_1v_1=\id_{P\p_1}+p\p_0\chi\p_1+\chi\p_0p\p_1\,,\nonumber
\eeqn
for some $\chi_i:P_i\rightarrow P_{i+1}$, $\chi\p_i:P\p_i\rightarrow P\p_{i+1}$. 

In this language, the class of trivial boundary conditions mentioned above can be represented by
the rank-one matrix factorisations 
\beq\label{trivialmf}
T:\quad P_1=\CC[X_i]\overset{p_1=1}{\underset{p_0=W}{\rightleftarrows}} \CC[X_i]=P_0\,.
\eeq
Any trivial matrix 
factorisation is equivalent to this special representative.

As mentioned above, the topological boundary degrees of freedom are
described by the cohomology of the boundary BRST-operator. In terms of
matrix factorisations the boundary BRST-operator on the boundary condition
changing sector between boundary conditions defined by matrix
factorisations $P$ and $P\p$ (topological open strings between the
D-branes associated to $P$ and $P\p$) is given by the graded commutator
with $Q_{\rm bd}$ on the space $\Hom_{\CC[X_i]}(P_1\oplus P_0,P_1\p\oplus
P_0\p)$ of boundary changing fields. More precisely, this operator acts on a boundary condition
changing field $\varphi\in\Hom_{\CC[X_i]}(P_1\oplus P_0,P_1\p\oplus
P_0\p)$ by
\beq
\varphi\mapsto Q\p_{\rm bd}\varphi-\sigma\p\varphi\sigma Q_{\rm bd}\,,
\eeq
where $\sigma=\id_{P_0}-\id_{P_1}$ is the grading operator on $P$. 
Since the BRST 
operator respects the grading, also its cohomology
$\HH(P,P\p)=\HH^0(P,P\p)\oplus\HH^1(P,P\p)$ is graded.

In the following, we will mostly be interested in the case that $W$ is
quasi-homogeneous, \ie $W(\lambda^{q_i}X_i)=\lambda^q W(X_i)$ for some
weights $q_i$, $q$, because these superpotentials directly correspond to
the superconformal field theories in the IR\footnote{Although 
  superpotentials are not renormalised, because of field redefinitions 
non-quasi-homogeneous superpotentials effectively flow to
quasi-homogeneous ones under the RG action.}. For such $W$, one can
also consider
quasi-homogeneous matrix factorisations, \ie matrix
factorisations 
\beq
P:\quad P_1\overset{p_1}{\underset{p_0}{\rightleftarrows}}P_0\,.
\eeq
together with representations $\rho_i$ of $\CC^*$ on the modules $P_i$ which are compatible with
the $\CC[X_i]$-action, such that the maps $p_i$ are
quasi-homogeneous:
\beq
\rho_0(\lambda)p_1\rho_1^{-1}(\lambda)=\lambda^{q\p} p_1\,,\quad
\rho_1(\lambda)p_0\rho_0^{-1}(\lambda)=\lambda^{q-q\p} p_0\,\quad{\rm
  for\; some}\; q\p\,.
\eeq
In the same way as quasi-homogeneous superpotentials correspond to 
conformal field theories in the IR, quasi-homogeneous matrix
factorisations correspond to conformal boundary conditions in these
CFTs, whereas matrix factorisations which are not quasi-homogeneous 
undergo an effective RG-flow\footnote{More details about this can be found in
\cite{Walcher:2004tx}.}. We will be mostly interested in
quasi-homogeneous matrix factorisations. 
\subsection{B-type defects and matrix factorisations}\label{sectdefectmf}
We will now consider the situation where a Landau-Ginzburg theory with
chiral superfields $X_i$ and a superpotential $W_1(X_i)$
is defined on the upper half plane, and a different Landau-Ginzburg theory with 
superfields $Y_i$ and a superpotential $W_2(Y_i)$ is defined on the lower half plane.
The two are separated by a defect on the real line. We would like to
describe those defects, which preserve
B-type supersymmetry. For this we will indeed follow the same strategy
used for the characterisation of B-type boundary conditions in
Landau-Ginzburg models reviewed in Section \ref{sectmfbd} above.

Again, only B-type supersymmetry preserves the B-type defect line.
Exactly as in the boundary case, the B-type supersymmetry
variation of the action of the theory on the UHP leads to a boundary term
(\ref{bulkvar}). The theory on the LHP gives a 
similar contribution, which however, because of the different relative
orientations of the boundary, has opposite sign. 
Therefore, the total B-type supersymmetry variation of
the action of the first Landau-Ginzburg model on the UHP and the
second one on the LHP is given by
\beqn\label{defectbulkvar}
\delta_B S&=& \delta_B S_D+\delta_B S_F\nonumber\\
\delta_B S_F&=&i\int dx^0 d\theta \big( \bar\epsilon (W_1-W_2) - \epsilon 
(\bar{W}_1 - \bar{W}_2 ) \big) .
\eeqn
Just as in the case of boundaries, $\delta_B S_D$ can be compensated by
an appropriate boundary term and $\delta_B S_F$ can be cancelled by 
introducing additional fermionic degrees of freedom on the defect. The same reasoning
as outlined in Section \ref{sectmfbd} for the case of boundary
conditions leads to the conclusion that B-type defects between the
two Landau-Ginzburg models are characterised by matrix factorisations
of the difference $W=W_1-W_2$ of the respective
superpotentials. As in 
the boundary case such a matrix factorisation gives rise to a defect
contribution $Q_{\rm def}$ to the BRST-charge, which squares to $W$.
Also the discussion of equivalence and trivial matrix factorisations
carries over directly from the discussion of boundary conditions.
Moreover, given two matrix factorisations
$P$, $P\p$ of $W$, in the same way as for boundary conditions, the cohomology
$\HH(P,P\p)$ of the BRST-operator induced by $Q_{\rm def}$ represents the
space of topological defect changing fields (topological closed
strings twisted by the two defects). Note however that defects carry
more structure than boundary conditions. Unlike boundaries, defects can
form junctions where more than two defects meet, 
and there are fields localised on these
junctions (topological closed strings twisted by more than two
defects). 
We will come back to this point later.

Before discussing more of the structure of defects in
Landau-Ginzburg models, we would like to remark that the conclusion that B-type defects
between two Landau-Ginzburg models are characterised by matrix
factorisation of the difference of their superpotentials is indeed
consistent with the folding trick. As alluded to in Section
\ref{sectgeneral}, the folding trick relates defects between two
two-dimensional theories ${\mathcal C}_1$ and ${\mathcal
  C}_2$ to boundary conditions in the product theory ${\mathcal
  C}_1\otimes\overline{\mathcal C}_2$, where $\overline{\mathcal C}_2$
is the theory ${\mathcal C}_2$ with left and right moving sectors
interchanged. Folding the Landau-Ginzburg model with superpotential
$W_2$ from the LHP to the UHP maps $x^\pm \mapsto x^\mp$ and likewise
$\theta^\pm \mapsto \theta^\mp$. In particular, it maps the D-term of the
theory on the LHP to the corresponding D-term on the UHP, while the
F-term changes sign\footnote{The measure $d^4\theta$ appearing in the
D-term is parity invariant, whereas the measure $d\theta^+ d\theta^-$
in the integral over chiral superspace changes sign.}
\beqa
S_{LHP}&=&
\int_{-\infty}^0 dx \int_{-\infty}^{\infty} dt d^4 \theta K(Y_i,\bar{Y}_i) + 
\int_{-\infty}^0 dx \int_{-\infty}^{\infty} dt 
\int d\theta^+ d\theta^- W_2(Y_i)  \\ \nn
&\mapsto&  
\int_{0}^{\infty} dx \int_{-\infty}^{\infty} dt d^4 \theta K(Y_i,\bar{Y}_i) - 
\int_{0}^{\infty} dx \int_{-\infty}^{\infty} dt 
\int d\theta^+ d\theta^- W_2(Y_i) \ .
\eeqa
The theory on the UHP obtained after folding up the
$W_2$-Landau-Ginzburg model from the LHP is the Landau-Ginzburg model
with chiral superfields $X_i$ and $Y_i$ whose 
K\"ahler potential is just the sum of the K\"ahler potentials of the
individual models, while its superpotential is the difference
$W=W_1-W_2$ of their superpotentials. As discussed in Section \ref{sectmfbd},
B-type boundary conditions in this model are indeed characterised by
matrix factorisations of $W$, which according to the folding trick
then carries over to B-type defects between Landau-Ginzburg models
with superpotentials $W_1$ and $W_2$. Thus, the folding trick provides
an alternative derivation for the fact that B-type defects between
Landau-Ginzburg models can be described by matrix factorisations of
the difference of their superpotentials. 

Note that if $W_1$ and $W_2$ are quasi-homogeneous with
respect to some $\CC^*$-action, then so is $W_1(X_i)-W_2(Y_i)$. 
As for boundary conditions, the corresponding 
quasi-homogeneous matrix factorisations give rise to conformal
defects in the IR CFT.
\section{Defect operation in Landau-Ginzburg models} \label{secoperation}

\FIGURE[r]{
\includegraphics[width=66mm]{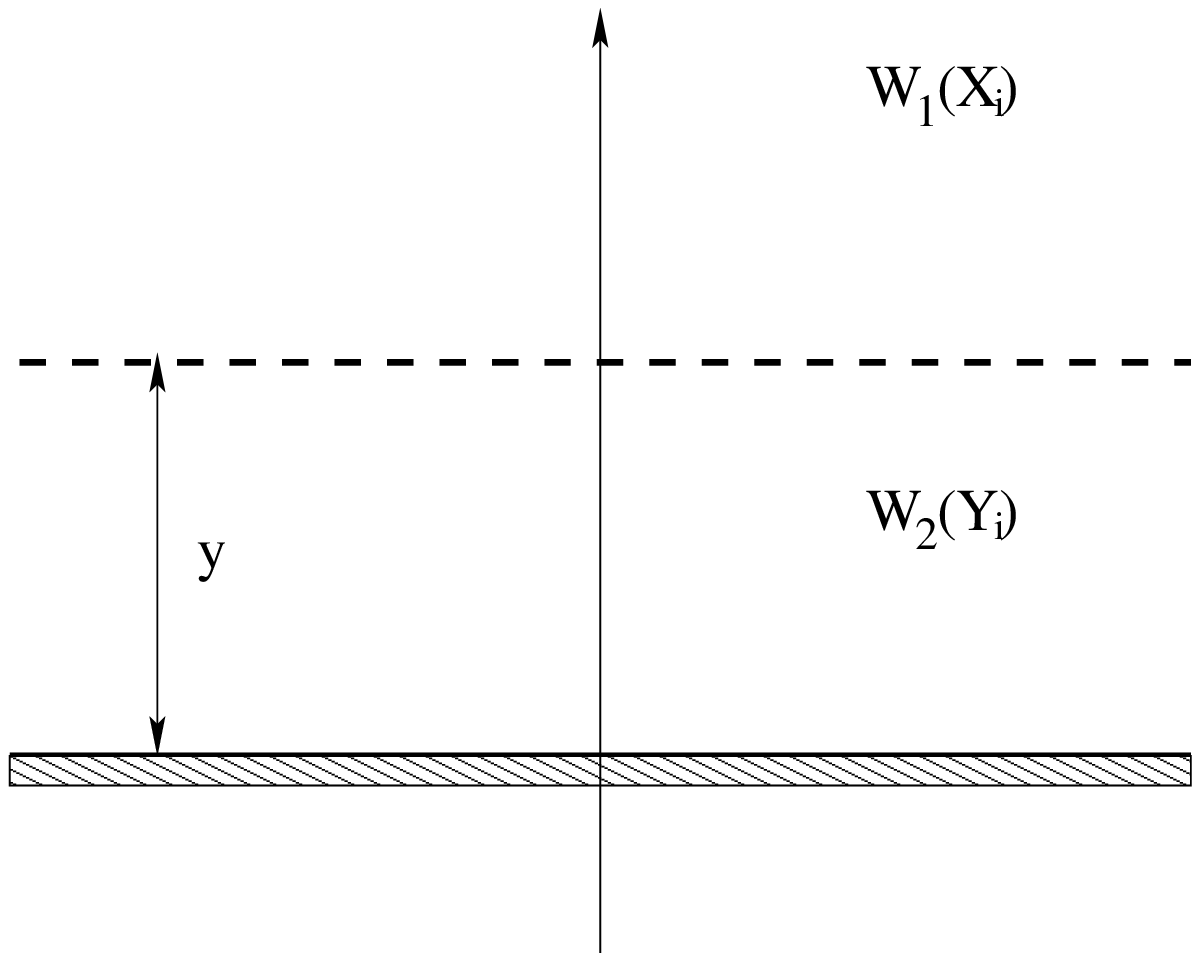}
\caption{\label{Fdefectbound}
Landau-Ginzburg models with superpotentials $W_1$ and $W_2$ on the
upper half plane separated by a defect (dashed line). Taking the defect to the
boundary ($y\to 0$), one obtains a new boundary condition on the
real line.}
}

\noindent
Having identified B-type defects between 
Landau-Ginzburg models as matrix factorisations, one can make use
of this rather elegant description to study properties of these
defects. For instance, one can investigate situations in which both
defects and boundaries, or in which various defects are present.

As discussed in Section \ref{sectgeneral}, upon topological twisting,
defects preserving the ap{\-}propriate supersymmetries become topological. This
means that correlation functions in the presence of such defects in the
topologically twisted theory do not change when the defects are
shifted on the world sheet. In particular, one can bring defects on top of each other or onto
world sheet boundaries. Note that for generic (even supersymmetric) defects in the
untwisted theory this is not possible. Only purely transmissive
defects can be shifted in the untwisted models, and correlation
functions
diverge when two non-topological defects approach each other, or such
a defect approaches a world sheet boundary. 
These singularities however disappear
upon topological twisting. Therefore, in the B-twisted
theory one can bring two B-type defects together to obtain another
one, and one can bring a B-type defect on 
top of a boundary satisfying B-type boundary conditions to obtain
a new boundary condition. That means, 
B-type defects can be
composed and act on B-type boundary conditions in the twisted models.
It is this composition
of B-type defects in Landau-Ginzburg models 
and their action on B-type boundary conditions which
we would like to describe in this section.

\subsection{Composition of defects and action on boundary
  conditions}\label{sectaction}

Let us start with the action of defects on boundary conditions. For
this consider a theory on the upper half plane consisting of a Landau-Ginz{\-}burg
model with chiral superfields $X_i$ and superpotential $W_1(X_i)$
defined on the domain $\RR+i\RR^{>y}$, and a Landau-Ginzburg model with
chiral superfields $Y_i$ and superpotential $W_2(Y_i)$ on the domain
$\RR+iy$ (\cf Figure \ref{Fdefectbound}). 
The two domains are separated by a B-type defect at
$\RR+iy$ defined by a matrix factorisation of
$W(X_i,Y_i)=W_1(X_i)-W_2(Y_i)$, and we impose B-type boundary
conditions on $\RR$ specified by a matrix factorisation of $W_2(Y_i)$.
Let us denote the respective defect and boundary BRST-charges by
$Q_{\rm def}$ and $Q_{\rm bd}$. They satisfy $Q_{\rm
  def}^2=(W_1-W_2)$, $Q_{\rm bd}^2=W_2$. What happens when the defect
is moved onto the boundary, \ie when $y\to 0$ is that in the
limit, both defect and boundary fermions $\pi_i^{\rm def}$, $\bar\pi_i^{\rm
  def}$ and $\pi_i^{\rm bd}$, $\bar\pi_i^{\rm bd}$ together with
$Q_{\rm def}$ and $Q_{\rm bd}$ are now defined on the world sheet
boundary $\RR$. The new 
boundary condition on $\RR$ created by moving the defect on top of the
original boundary condition has boundary BRST-charge 
\beq
Q\p_{\rm bd}=Q_{\rm def}+Q_{\rm bd}\,.
\eeq
Since $Q_{\rm bd}$ and $Q_{\rm def}$ anti-commute, 
\beq
\left(Q\p_{\rm bd}\right)^2=Q_{\rm def}^2+Q_{\rm
  bd}^2=W_1(X_i)-W_2(Y_i)+W_2(Y_i)=W_1(X_i)
\eeq
and therefore $Q'_{\rm bd}$ is indeed a 
BRST-charge of a B-type boundary condition in a
Landau-Ginzburg model with superpotential $W_1$.
Note however that $Q\p_{\rm bd}$ still involves the chiral superfields
$Y_i$ of the Landau-Ginzburg model squeezed in between defect and
boundary. In the limit, they are promoted to new boundary degrees of
freedom.

In terms of matrix factorisations this can be formulated as follows.
Let
\beq\label{defmf}
P:\quad P_1\overset{p_1}{\underset{p_0}{\rightleftarrows}} P_0\,,\qquad
p_1p_0=(W_1(X_i)-W_2(Y_i))\id_{P_0}\,,\quad 
p_0p_1=(W_1(X_i)-W_2(Y_i))\id_{P_1}
\eeq
be the matrix factorisation of $W(X_i,Y_i)=W_1(X_i)-W_2(Y_i)$
representing the defect at $\RR+iy$, and let the original boundary
condition on $\RR$ correspond to the matrix factorisation 
\beq\label{bcmf}
Q:\quad Q_1\overset{q_1}{\underset{q_0}{\rightleftarrows}} Q_0\,,\qquad
q_1q_0=W_2(Y_i)\id_{Q_0}\,,\quad 
q_0q_1=W_2(Y_i)\id_{Q_1}
\eeq
of $W_2(Y_i)$. The new boundary condition arising on $\RR$ in the
limit $y\to 0$ is given by the {\it tensor product matrix
  factorisation}
\beqn\label{defectbctp}
Q\p:&&Q\p_1=\left(P_1\otimes_{\CC[Y_i]} Q_0\right)\oplus \left(P_0\otimes_{\CC[Y_i]} Q_1\right)
\overset{q\p_1}{\underset{q\p_0}{\rightleftarrows}}
Q\p_0=\left(P_0\otimes_{\CC[Y_i]}Q_0\right)\oplus \left(P_1\otimes_{\CC[Y_i]}Q_1\right)\nonumber\\
&&{\rm with}\qquad q\p_1=\left(\begin{array}{cc} p_1 & -q_1 \\ q_0 &
p_0\end{array}\right)\,,\qquad
q\p_0=\left(\begin{array}{cc} p_0 & q_1 \\ -q_0 &
p_1\end{array}\right)\,.
\eeqn
Since $Q\p$ represents a B-type boundary condition in the Landau-Ginzburg model
with chiral superfields $X_i$ and superpotential $W_1(X_i)$, it
has to be regarded as a matrix factorisation over $\CC[X_i]$. However,
by construction, the $Q\p_i$ are really free $\CC[X_i,Y_i]$-modules, 
therefore in particular free $\CC[X_i]$-modules of infinite rank,
which means that the matrix factorisation $Q\p$ defined by
\eq{defectbctp} is a matrix factorisation of infinite rank over
$\CC[X_i]$.

Thus,
moving a B-type defect on top of a B-type boundary, one obtains a boundary
condition defined by a matrix factorisation of infinite rank. This is
due to the new boundary degrees of freedom arising from the bulk fields
$Y_i$ of the Landau-Ginzburg model squeezed in between boundary and defect.

As it turns out, this is only an artifact of the
construction. The matrix factorisations \eq{defectbctp}
obtained from finite rank matrix factorisations $P$ and
$Q$ are always equivalent up to trivial matrix factorisations to
finite rank matrix factorisations of $W_1(X_i)$. That 
$Q\p=P\otimes Q$ 
is of infinite rank is entirely due to the appearance of spurious trivial matrix
factorisations (brane-anti-brane pairs) which are physically
irrelevant. Extracting the reduced finite rank matrix factorisation
from it is the non-trivial part of the analysis of the
action of B-type defects on B-type boundary conditions in topological 
Landau-Ginzburg models. In Section \ref{sectdefmf} below we will
present an argument why the matrix factorisations $Q\p$ can always be
reduced to finite rank, and we will discuss a method to extract
the reduced matrix factorisations. Explicit examples will be analysed 
in Section \ref{comparison}.

\FIGURE[ht]{
\includegraphics[width=120mm]{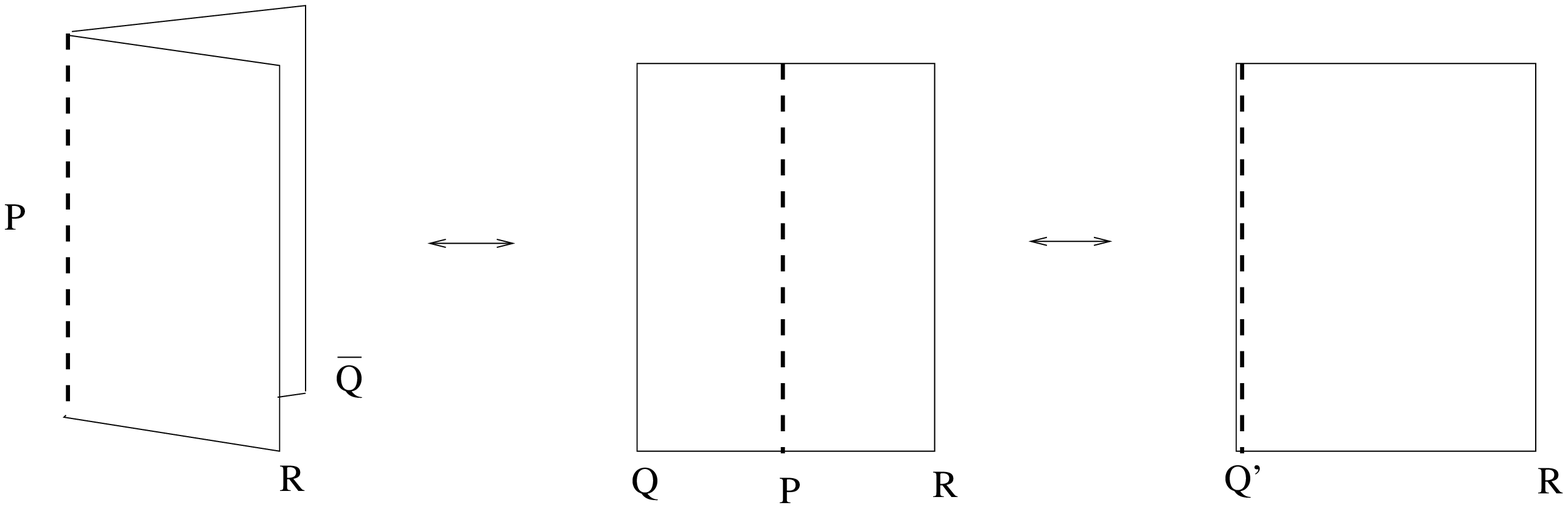}
\caption{
From left to right: 1) 
The configuration with defects $P$ and $P'=\bQ \otimes R$,
2) open strings between boundary conditions $R$ and $Q$, 
twisted by the defect $P$,
3) As a limit of 2) one obtains open strings between a new boundary
condition $Q'= Q\otimes P$ and $R$.
}
}
Before turning to a discussion of the composition of B-type defects,
let us remark that the representation of the action of B-type defects
on B-type boundary conditions in terms of the tensor product
\eq{defectbctp} is also very natural from the point of view of the
topological spectra. As explained in Section \ref{sectdefectmf} the
topological defect changing spectra between two B-type defects
represented by matrix factorisations $P\p$ and $P$ of $W=W_1-W_2$ is
given by the BRST-cohomology 
$\HH^*(P\p,P)$.

Now, for $P\p$ one can
in particular choose a tensor product $P\p=R\otimes \bar Q$ of
matrix factorisations $R$ of $W_1$ and $\bar Q$ of $-W_2$. (Given a
matrix factorisation $Q$ of $W_2$, we denote by $\bar Q$ the matrix
factorisation of $-W_2$ obtained by $q_1\mapsto -q_1$.) Such a tensor
product matrix factorisation in fact represents a purely reflexive
defect, \ie a tensor product of boundary conditions in the two
Landau-Ginzburg models. Therefore, the topological cylinder amplitude
with defects corresponding to $P\p$ and $P$ inserted along the
cylinder, is in fact nothing but the topological amplitude on a
strip with boundary conditions corresponding to $Q$ and $R$ along the
boundaries, and a defect corresponding to $P$ inserted between them. Hence, for
$P\p=R\otimes\bar Q$ the spectrum $\HH^*(P\p,P)$ in fact also
represents the spectrum of topological open strings between the
D-branes corresponding to $R$ and $Q$ twisted by the defect
corresponding to $P$. Moreover, since in the topologically twisted theory
B-type defects are topological, the spectrum should not change when
moving the defect. In particular, it should not change, when
bringing the defect on top of one of the boundaries, the one
corresponding to $Q$ say. From this it
follows that the spectrum $\HH^*(P\p,P)$  should indeed also describe
the spectrum of topological open strings between the D-brane
corresponding to $R$ on one side and the D-brane described by $Q\p$ 
arising from bringing the defect $P$ onto the boundary condition $Q$
on the other: 
\beq
\HH^*(P\p=R\otimes\bar Q,P)\cong\HH^*(R,Q\p)\,.
\eeq
But from the construction of the BRST-operator, it is easy to see that
indeed
\beq\label{brstshift}
\HH^*(R\otimes\bar Q,P)\cong\HH^*(R,P\otimes Q)\,,
\eeq
which shows that the tensor product factorisation $Q\p=P\otimes
Q$ has the spectrum expected from the matrix factorisation
representing the boundary condition obtained by moving the defect
described by $P$ onto the boundary with boundary condition $Q$.

Completely analogously to the action of $B$-type defects on $B$-type boundary
conditions one can describe the action of them on other $B$-type
defects, \ie their composition.
On the level of matrix factorisation, the latter is also 
represented by taking the tensor product of the matrix factorisations
representing the defects which are being composed.
For this replace in the discussion above
the matrix factorisation (\ref{bcmf}) by a matrix factorisation of
$W_2(Y_i)-W_3(Z_i)$ over $\CC[Y_i,Z_i]$ representing a defect between the
Landau-Ginzburg models with superpotentials $W_2(Y_i)$ and 
$W_3(Z_i)$ respectively. The tensor product (\ref{defectbctp}) then
gives rise to an infinite rank matrix factorisation of
$W_1(X_i)-W_3(Z_i)$ representing the defect emerging as the
composition of the two defects. As in the case of the action on boundary
conditions it is in fact equivalent modulo trivial matrix
factorisations to a finite rank one, and also for the analysis of the
composition of defects the challenge lies in the reduction of the
infinite rank tensor product matrix factorisation to finite rank.

We would like to close the general discussion of composition of B-type
defects in Landau-Ginzburg models and their action on B-type boundary
conditions with the following remark.
As pointed out in Section \ref{sectdefectmf}, one fundamental
difference between boundary conditions and defects is the possibility
of the latter to form junctions, which also carry fields. The
discussion above in fact suggests a simple method to calculate the
topological spectra of fields localised on the junctions formed by $n$
B-type defects in Landau-Ginzburg models. Namely, let $W_1,\ldots,
W_n$, $W_{n+1}=W_1$
be superpotentials, and for $1\leq i \leq
n$ let $P^i$ be matrix factorisations of $W_i-W_{i+1}$
representing B-type defects between the respective Landau-Ginzburg
models. To calculate the topological spectrum
$\HH^*(P^1,\ldots,P^n)$ of fields on the
junction formed by these defects (topological closed strings twisted
by all of them) we note that as above, the topological spectra should
not change when shifting the defects. So in particular, we can
bring the last $n-1$ of them on top of each other, and the spectrum of fields
on the junction is identical to the spectrum of defect changing
fields between the defect represented by $P^1$ and the one obtained by
composing the defects associated to $P^i$ with $i>1$. Since the latter
is represented by the matrix factorisation $P^2\otimes\ldots\otimes
P^n$ one obtains
\beq\label{defcomposition}
\HH^*(P^1,\ldots,P^n)\cong\HH^*(\bar P^1,P^2\otimes\ldots\otimes
P^n)\,.
\eeq
\subsection{Defect operation and matrix factorisations}\label{sectdefmf}
In the Section \ref{sectdefectmf} we have argued that similarly to B-type boundary
conditions, also B-type defects in Landau-Ginzburg models can be
described by means of matrix factorisations. We have explained that
in this formulation, 
the action of these defects on B-type boundary conditions and defects
has a simple realisation in terms of the tensor product
\eq{defectbctp} of the respective matrix factorisations.
As was pointed out in the previous section, the tensor product
matrix factorisations obtained in this way are of infinite rank however.
Here we will argue that they indeed are
always equivalent to matrix
factorisations of finite rank. That means, it is always possible to
reduce them to matrix factorisations of finite rank by splitting off
infinitely many trivial matrix factorisations. It is this reduction
which is the non-trivial part in the analysis of the action of B-type
defects, and we will discuss a method to deal with it below. We will
focus on the action of B-type defects on B-type
boundary conditions, but the discussion of the composition of 
defects works exactly analogously.

The basic idea we employ to show that the tensor product matrix
factorisations obtained are equivalent to finite rank factorisations
is to identify the reduced rank as the dimension of a certain
BRST-cohomology group, which can be calculated directly from the
infinite rank representative. (In a geometric context, one would want to count the
bosonic open strings between the D-brane under consideration and the basic D-brane with Neumann
boundary conditions in all directions, carrying only one type of charge.)
This argument only works in the case that $W$ and the matrix
factorisations under consideration are quasi-homogeneous\footnote{Indeed it also works for
non-homogeneous matrix factorisations defined over Laurent rings
instead of polynomial rings.}.
Since we are mostly interested in
quasi-homogeneous matrix factorisations, we will restrict the discussion to
this case, but we believe that the statement also holds in the general
situation.

Let us start the discussion by the following remark. Consider a matrix
factorisation
\beq
Q:\quad Q_1\overset{q_1}{\underset{q_0}{\rightleftarrows}} Q_0\,,\qquad
q_1q_0=W(X_i)\id_{Q_0}\,,\quad 
q_0q_1=W(X_i)\id_{Q_1}
\eeq
of $W(X_i)$ over the ring $\CC[X_i]$. Now suppose that $q_1$ or $q_0$
have an entry which is a unit (\ie an invertible element) in
$\CC[X_i]$. It is easy to see
that in this case there is an equivalence 
$(u_i,v_i=u_i^{-1})$ as in \eq{Mequiv} which brings $Q$ into the form
$Q\cong Q\p\oplus T$, where $T$ is the trivial matrix factorisation
\eq{trivialmf}. In particular, $Q$ can be reduced to $Q\cong Q\p$.
This can be done until there are no more unit
entries in the matrix factorisation, in which case no trivial matrix
factorisation can be split off anymore in this way. Let us assume this
to be true for the matrix factorisation $Q$ and let $N$ be its
rank. Under these circumstances, the rank of 
$Q$ can be calculated as the dimension of the BRST-cohomology $\HH^0(Q,S)$
\beq\label{rank}
{\rm rank}(Q)=\dim\HH^0(Q,S)\,,
\eeq
where $S$ is the tensor product\footnote{As in the definition of $D_g$
in \eq{gmf} below, the rank-one factors are not necessarily unique, but the
resulting tensor product matrix factorisation is up to equivalence.} of the rank-one factorisations $P^i$
defined by $p^i_1=X_i$, $p^i_0=B_i(X_i)$ with $W=\sum_i X_i B_i$.
This can be seen as follows. The Koszul resolution of the module
$M=\CC[X_i]/(X_i)$ regarded as a $\CC[X_i]$-module can be used to
construct an 
$R:=\CC[X_i]/(W)$-free resolution of $M$, which after $l$ steps 
turns into the two-periodic resolution of
$\coker(p_1)$ defined by $p_1$ and $p_0$. Here, $l=n$ is the number of
variables $X_i$ if $n$ is even, and $l=n-1$ if $n$ is odd. 
Thus,
\beq
\Ext^i_R(\cdot,\coker(p_1))\cong\Ext^{i+l}_R(\cdot,M)\,.
\eeq
This has been discussed in detail in Section 4.3 of \cite{Enger:2005jk}. 
Using the general fact \cite{Enger:2005jk} that for all $i>0$
\beq
\HH^0(P,Q)\cong\Ext^{2i}_R(\coker(p_1),\coker(q_1))\,,\;
\HH^1(P,Q)\cong\Ext^{2i-1}_R(\coker(p_1),\coker(q_1))\,,\nonumber
\eeq
one obtains
\beq\label{exts}
\HH^0(Q,S)\cong \Ext_R^{2+l}(\coker(q_1),M)\,.
\eeq
This $\Ext$-group can be calculated by means of the two-periodic
resolution 
\beq\label{res}
\ldots\stackrel{q_1}{\longrightarrow}R^N\stackrel{q_0}{\longrightarrow}R^N\stackrel{q_1}{\longrightarrow}R^N
\longrightarrow\coker(q_1)\longrightarrow 0\,.
\eeq
Namely, it is given by the cohomology of the complex obtained
by applying the functor $\Hom_R(\cdot,M)$ to the resolution \eq{res}. Since the $q_i$ (in a certain
basis) only have
non-unit homogeneous entries, the differentials of this complex all vanish, and therefore
the cohomology in every degree is given by $M^N$. In particular, the
dimension of the cohomology groups is $N$, the rank of $Q$. 

But now, $\dim\HH^0(Q,S)$ does not change when one adds trivial
matrix factorisations to $Q$. This implies that $\dim\HH^0(Q,S)$
indeed calculates the reduced rank of the matrix factorisation $Q$,
\ie the rank of the matrix factorisation obtained from $Q$ by
splitting off all trivial matrix factorisations in the way described
above\footnote{A priori it might be possible that $Q$ is equivalent to a
matrix factorisation with even smaller rank.}.

We will use this to show that the reduced rank of tensor product
matrix factorisations representing the boundary conditions obtained by
applying a B-type defect to a B-type boundary condition is always
finite (assuming that the factor matrix factorisations are of finite
rank). 
So let $P$ as in \eq{defmf} represent a B-type defect between
two Landau-Ginzburg models, $Q$ as in \eq{bcmf} a B-type boundary
condition in one of them, and $Q\p$ defined in
\eq{defectbctp} their tensor
product. To show that the reduced rank of $Q\p$ is finite we again
make use of \eq{exts}. As above we use the resolution \eq{res} for
$Q\p$ to compute the $\Ext$-groups, which are then given by the cohomology
of the sequence
\beq
\ldots\stackrel{\wt q\p_0}{\longrightarrow}
(M\p)^{N\p}\stackrel{\wt q\p_1}{\longrightarrow}
(M\p)^{N\p}\stackrel{\wt q\p_0}{\longrightarrow}
(M\p)^{N\p}\stackrel{\wt q\p_1}{\longrightarrow}\ldots\,
\eeq
Here $N\p$ denotes the rank of $Q\p$, $M\p=\CC[X_i,Y_i]/(X_i)$ and
$\wt q\p_i$ are obtained from the $q\p_i$ by setting $X_i=0$.
But similarly as in the discussion of \eq{brstshift}, one recognises
this complex as the one computing $\HH(Q,\overline{\wt P})$,
where $\wt P$ is the matrix factorisation over $\CC[Y_i]$ 
obtained from $P$ by setting $X_i=0$. It is in particular a finite
rank matrix factorisation of $W_2(Y_i)$. We therefore obtain
\beq
\HH^0(Q\p=P\otimes Q,S)\cong\HH^i(Q,\overline{\wt P})\,,
\eeq
for some $i$, 
where the latter is the BRST-cohomology between two finite rank matrix
factorisations of $W_2(Y_i)$, which in particular is finite
dimensional. Hence the reduced rank of $Q\p$ is finite.

Having established that $Q\p$ can be reduced to a matrix factorisation of finite
rank, we would now like to comment on how to obtain a reduced
form. To find the explicit equivalence on the level of matrix factorisations is difficult
in general. (A very simple example is discussed in Appendix \ref{explicitequiv}.) 
On the level of modules, what
one has to do is to regard $\coker(q\p_1)$ as an
$R$-module and split off all free summands. A trick, which will prove
useful in the examples presented in Section \ref{comparison} is the
following. Instead of analysing $\coker(q\p_1)$ one can consider the module
$V=\coker(p_1\otimes\id_{Q_0},-\id_{P_0}\otimes q_1)$. This module has
the $R$-free resolution
\beq
\ldots\stackrel{q\p_1}{\longrightarrow}
Q\p_0\stackrel{q\p_0}{\longrightarrow}
Q\p_1\stackrel{q\p_1}{\longrightarrow}
Q\p_0\stackrel{q\p_0}{\longrightarrow}
Q\p_1
\stackrel{(p_1\otimes\id_{Q_0},-\id_{P_0}\otimes
q_1)}{\longrightarrow} 
P_0\otimes Q_0\longrightarrow V\longrightarrow 0,
\eeq
which after two steps turns into the $R$-free resolution of
$\coker(q\p_1)$ obtained from the 
matrix factorisation $Q\p$. 
Therefore, instead of reducing $\coker(q\p_1)$, we can just as well
reduce $V$ and take the matrix factorisation which can be obtained
from a resolution of the reduced module by chopping off the first two terms. 
Indeed, in the examples presented in Section \ref{comparison} this trick
will prove to be very useful, because $V$ itself will already be of
finite rank. 

Completely analogously to the action of $B$-type defects on $B$-type boundary
conditions, the composition of $B$-type defects can be described.
For this replace
the matrix factorisation (\ref{bcmf}) with a matrix factorisation of
$W_2(Y_i)-W_3(Z_i)$ over $\CC[Y_i,Z_i]$ representing a defect between the
Landau-Ginzburg model with superpotential $W_2(Y_i)$ and the one with
superpotential $W_3(Z_i)$. The tensor product (\ref{defectbctp}) then
gives rise to an infinite rank matrix factorisation of
$W_1(X_i)-W_3(Z_i)$, which as in the case of the action on boundary
conditions is in fact equivalent to a finite dimensional one. Thus,
from $B$-type defects between Landau-Ginzburg theories with
superpotentials $W_1(X_i)$ and $W_2(Y_i)$, and $W_2(Y_i)$ and $W_3(Z_i)$
respectively, one obtains one between the Landau-Ginzburg theories
with superpotentials $W_1(X_i)$ and $W_3(Z_i)$. 
\section{Symmetry defects}\label{sectgrouplike}
If a two-dimensional field theory exhibits symmetries, \ie 
automorphisms of its Hilbert space which commute with energy and
momentum operators, then these 
give rise to topological defects. The corresponding
defect
operators are simply given by the automorphisms themselves, 
and the closed string sectors
twisted by such defects are the ordinary twisted sectors known from
orbifold constructions\footnote{In fact, also dualities between different
  theories can give rise to such defects.}. Obviously, these defects compose according to
the symmetry group of the theory, and in particular every such defect
has an inverse. They are group-like defects as discussed in
\cite{Frohlich:2006ch}.

For a Landau-Ginzburg model with chiral superfields $X_i$ and
superpotential $W$ there is a simple class of symmetries,
whose action is defined by linear and 
unitary\footnote{The standard K\"ahler potential $K=\sum_i\bar X_i
  X_i$ has to be invariant.} 
action on the superfields $X_i$:
\beq\label{gaction}
X_i\mapsto g(X_i)\,,\quad {\rm such}\,{\rm that}\,\,W(g(X_i))=W(X_i)\,.
\eeq
We will denote the group of these transformations by $\Gamma$. 
The corresponding defects can easily be described by means of gluing
conditions of the chiral superfields along the defect. Let us
consider a Landau-Ginzburg model with superpotential $W$ on the full
plane with such a defect along the real line. Denote by $X_i$ and
$Y_i$ the chiral superfields on the UHP and the LHP respectively.
Then for every $g\in\Gamma$ as above, one can define a defect $D_g$ by
imposing the gluing conditions
\beq\label{glcond}
\left(X_i(x+iy)-g(Y_i)(x-iy)\right)\to 0\quad{\rm for}\quad y\to 0
\eeq
on the chiral superfields on the UHP and LHP along the real line.
Obviously, these gluing conditions cancel the supersymmetry variation 
\eq{defectbulkvar} of the bulk F-term in the presence of the defect
without the introduction of any additional degrees of freedom.
Moreover, these defects are also compatible with A-type supersymmetry. 
To see this, consider the full supersymmetry variation (\ref{delta})
of the F-term of the theory on the upper half plane. The result is
\beq
\delta S= i \int_{\RR} dx^0 \big( \bar{\epsilon}_+ \omega_+^{(1)} - \bar{\epsilon}_-
\omega_-^{(1)} + \epsilon_- \bar{\omega}_-^{(1)} - 
\epsilon_+ \bar{\omega_+}^{(1)} \big).
\eeq
where we have expanded the chiral superfield $W_1$ as
\beq
W_1= w^{(1)}(y^\pm) +  \theta^\alpha\omega_\alpha^{(1)} (y^\pm) + 
\theta^+\theta^- F^{(1)}(y^\pm),
\eeq
This can be compensated by the variation of a theory defined on the lower
half plane if
\beqa
(\omega_\pm^{(1)}(x+iy) &-& \omega_\pm^{(2)}(x-iy) ) \to 0
\ ,\\ \nn
(\bar{\omega}_\pm^{(1)}(x+iy) &-& \bar{\omega}_\pm^{(2)}(x-iy) ) 
\to 0 
\eeqa
modulo total derivatives in the limit $y\to 0$. 
As one easily checks, these conditions are 
satisfied in the case that the
chiral superfields obey the gluing relations (\ref{glcond}).
The latter furthermore imply gluing conditions
\beq
Q_\pm^{(1)} = Q_{\pm}^{(2)}, \quad \bQ_\pm^{(1)} = \bQ_{\pm}^{(2)}
\eeq
for the Landau-Ginzburg supercharges (\ref{lgsupercharges}) along the
defect line, which ensures that indeed 
the full $N=(2,2)$ supersymmetry is preserved.

Since the defects are topological, one can compose them with the
obvious result 
\beq
D_gD_{g\p}=D_{gg\p}\,.
\eeq
As mentioned above the defect spectra obtained
from defects defined by group actions are nothing but
the twisted spectra usually discussed in the context of the
corresponding orbifold models. 
We refer to
\cite{Vafa:1989xc,Intriligator:1990ua}, 
for a discussion of the twisted sectors in Landau-Ginzburg orbifolds.

Even though  these defects have a very nice and simple description
not involving  new degrees of freedom on the
defect, we would  like to make contact with the discussion 
of the previous sections and show how to formulate them in 
terms of matrix factorisations. 
Here, we can take inspiration from a similar discussion in the context of boundary
conditions. 
B-type boundary conditions for Landau-Ginzburg models had
first been introduced in \cite{Hori:2000ck,Hori:2000ic} without the
introduction of additional boundary degrees of freedom. After the
discovery that matrix factorisations provide more general boundary
conditions, it was proposed in
\cite{Govindarajan:2006uy,Ezhuthachan:2005jr} that the original
boundary conditions of \cite{Hori:2000ck,Hori:2000ic} can indeed be
realised as matrix factorisations, having one linear factor
representing the gluing conditions of the chiral fields along the
boundary. This suggests that the group
like defects discussed above should be realised as linear matrix
factorisations as well. Indeed, for every 
$g$ as above $W(X_i)-W(Y_i)$ can be factorised as\footnote{This
  factorisation may not be unique, but the matrix factorisations \eq{gmf} 
  resulting from different choices of $A_j$ in \eq{groupfactor} are equivalent.}
\beq\label{groupfactor}
W(X_i)-W(Y_i)=\sum_j(X_j-g(Y_j))A_j(X_i,Y_i)\,,
\eeq
for some polynomials $A_j(X_i,Y_i)$, generalising the prescription for $g=1$
in \cite{Kapustin:2004df}.
We propose that 
the defects $D_g$ can then be represented by the tensor product matrix
factorisations
\beq\label{gmf}
D_g= \bigotimes_i P^i
\eeq
of the rank-one factorisations defined by
\beq
p^i_1=(X_i-g(Y_i))\,,\quad p^i_0=A_i(X_i,Y_i)\,.
\eeq
Let us gather some evidence for this proposal. 
It is indeed very easy to verify that the matrix factorisations 
\eq{gmf} lead to the desired action on B-type defects and boundary
conditions. To see this, consider a matrix factorisation $D_g$ as
defined above and a matrix factorisation $Q$ which is either a matrix
factorisation of $W(Y_i)$ corresponding to a B-type boundary
condition, or a matrix factorisation of $W(Y_i)-W(Z_i)$ representing
another defect. We set $R:=\CC[X_i]/(W)$ in the first,
$R:=\CC[X_i,Z_i]/(W(X_i)-W(Z_i))$ in the second case.

The result of the action of $D_g$ on $Q$ is given by
the matrix factorisation $D_g\otimes Q$. To reduce this, we employ the
same trick used to get \eq{exts}. Namely, the module
\beq
M:=\coker((X_1-g(Y_1))\id_{Q_0},\ldots,(X_n-g(Y_n))\id_{Q_0},q_1)
\eeq
has an $R$-free resolution which after $l=n+1$ for $n$ odd, and $l=n$ for $n$
even steps turns into the matrix factorisation $D_g\otimes Q$. (This
resolution is related to the Koszul complex, and is discussed in a
similar context in Section 4.3 of \cite{Enger:2005jk}.)
Therefore the matrix factorisation $D_g\otimes Q$ is equivalent
to the matrix factorisation
into which the $R$-free resolution of $M$ turns  after $l$ steps.
But $M$ is nothing else than
\beq
M\cong\coker(q_1(Y_i=g^{-1}(X_i)))
\eeq
which obviously has a completely two-periodic resolution,
namely the matrix factorisation
$Q(Y_i=g^{-1}(X_i))$ over $R$. Thus, $D_g$ acts on matrix
factorisations by setting $Y_i=g^{-1}(X_i)$. In particular one obtains
the desired composition of the symmetry defects $D_g$, because
$D_g\otimes D_{g\p}$ is equivalent to $D_{gg\p}$. 

Also the analysis of defect spectra supports the identification of the matrix
factorisations \eq{gmf} with group like defects. The spectra
associated to the symmetry defects do indeed agree with the spectra of
bulk fields twisted by the respective group elements as calculated in 
\cite{Intriligator:1990ua,Vafa:1989xc}. More precisely, one can show
that the defect spectra $\HH^*(D_g,D_1)$ are isomorphic to the $g$-twisted bulk
Hilbert spaces\footnote{By means of the
composition of the $D_g$ discussed above
$\HH^*(D_{g\p},D_{g})\cong\HH^*(D_{g\p g^{-1}},D_1)$.}. 
For instance, in the case of a Landau-Ginzburg model with a single
chiral superfield $X$ 
it is indeed very easy to see by direct calculation 
that $\HH^*(D_1,D_1)$ is purely bosonic and isomorphic to the bulk chiral
ring $\CC[X]/(\partial W)$, \ie the untwisted bulk Hilbert space. 
For $g\neq 1$, on the other hand there are
no bosons in the defect spectra $\HH^*(D_g,D_1)$, and only a single
fermion $\omega$, corresponding to the unique ground state in the
$g$-twisted sector of the orbifold. 

This easily generalises to tensor
products of this situation (in particular the $g$ act diagonally on
the $X_i$), in which case each tensor factor
contributes to $\HH^*(D_g,D_1)$ either
polynomials in $\CC[X_i]/(\partial_i W)$ in case $X_i$ is
$g$-invariant, or a fermion $\omega_i$, if it is not. Hence,
$\HH^*(D_g,D_1)$ is spanned by polynomials in $g$-invariant variables
multiplied by one fermion for each variable which is not
$g$-invariant. This can be written as 
\beq
\HH^*(D_g,D_1)\cong\CC[X_i^{g-{\rm inv}}]/(\partial_i(W_{g-{\rm
inv}}))\prod_{g(X_j)\neq X_j}\omega_j\,,
\eeq
with $W_{g-{\rm inv}}$ the polynomial obtained from $W$ by setting
all non-$g$-invariant variables to zero. It is easily recognised as
the $g$-twisted orbifold sector obtained in 
\cite{Vafa:1989xc,Intriligator:1990ua}. 

Indeed, the statement that the defect spectra $\HH^*(D_g,D_1)$ are
isomorphic to the $g$-twisted sectors is true in the general
situation. The proof for the general case is presented in Appendix 
\ref{sectglspectra}\footnote{That the spaces $\HH^*(D_1,D_1)$ for
general $W$ coincide with the
bulk-chiral rings has also been observed in \cite{Kapustin:2004df}}.
\section{Defects in minimal models}\label{comparison}
Up to now we have discussed symmetry defects in arbitrary Landau-Ginzburg
models. The matrix factorisations describing these defects, their
action on B-type boundary conditions and their composition properties
have been discussed in Section \ref{sectgrouplike}. Here, we would
like to discuss more general defects in 
Landau-Ginzburg models with one
chiral superfield and superpotential $W(X)=X^d$, and in 
their closely related cousins, theories with one additional superfield
and superpotential $W(X)=X^d+Z^2$.
The bulk chiral rings of these two theories are equivalent,
but there are differences in the D-brane spectra, as discussed in
\cite{Kapustin:2003rc}. 
Indeed the two theories can be regarded as $\ZZ_2$ orbifolds of each
other, and therefore, adding a further square leads again to the initial
theory. On the level of matrix factorisation this property is known as
Kn\"orrer periodicity.

In the IR these models become respectively $N=2$ superconformal minimal models and
$\ZZ_2$-orbifolds thereof. Both these models share the same Hilbert
space, but differ in the action of $(-1)^F$. They are well understood conformal field
theories in which conformal defects can be explicitly studied. 

In \ref{LGapproach}, we will construct and
analyse defects within the Landau-Ginzburg framework presented in the
previous sections. 
In \ref{CFTapproach} we will make contact with the
CFT-analysis. We will restrict our attention to defects between one
and the same Landau-Ginzburg model, in which case constructions for
the corresponding conformal defects are known. The case of defects
between Landau-Ginzburg models with 
different superpotentials will be investigated in \cite{toappear}.
\subsection{Landau-Ginzburg approach}\label{LGapproach}
Let us start with the case $W(X)=X^d$. The model with superpotential
$W(X,Z)=X^d+Z^2$ will be discussed later in Subsection \ref{sectlgz2}.
There are certain obvious candidates for defect matrix factorisations of
$W(X)-W(Y)=X^d-Y^d$. On the one hand, these are the tensor product
matrix factorisations
\beq\label{deftpfact}
T_{i,j}:\quad t_1^{i,j}=\left(\begin{array}{cc} X^i & Y^j\\ Y^{d-j}
& X^{d-i}\end{array}\right)\,,\quad
t_0^{i,j}=\left(\begin{array}{cc} X^{d-i} & -Y^j\\ -Y^{d-j}
& X^{i}\end{array}\right)\,.
\eeq
On the other hand there are 
``permutation type'' matrix factorisations of the form
\beq\label{defpfact}
P^d_I: \quad p_1^I=\prod_{a\in I}(X-\eta^a Y)\,,\quad
p_0^I=\prod_{a\in \{0,\ldots,d-1\}-I}(X-\eta^a Y)\,,
\eeq
where $\eta$ is an elementary $d$th root of unity and $I$ is a strict
subset of $\{0,\ldots,d-1\}$. These defects generalise the symmetry
defects discussed in Section \ref{sectgrouplike}. Namely, the
Landau-Ginzburg model with superpotential $W=X^d$ allows for the
operation of a symmetry group $\ZZ_d$ on the superfield 
\beq
i\in\ZZ_d:\quad X\mapsto \eta^i X\,,
\eeq
and the corresponding defect matrix factorisations $D_i$ agree with
the matrix factorisations $P_{\{i\}}$ of \eq{defpfact}.
\subsubsection{Composition of permutation type matrix defects}
The action of the permutation type matrix factorisations $P_{\{i\}}$ 
has already been discussed in Section \ref{sectgrouplike}. Here we
would like to analyse the composition of defects represented by 
$P_I$ for arbitrary $I$. As
it will turn out, we will only have to analyse the action of 
$P_I$ with $|I|=2$, because successively composing such $P_I$ one can generate all
other $P_I$ as well. Considerations will be restricted to the case
where $I$ is a set of successive integers modulo $d$, because these
defects have a simple representation in the respective conformal field
theories.

As a warm up, let us consider the composition of two defects
corresponding to matrix factorisations $P_I$ with $|I|=2$,
$P_{\{m,m+1\}}$ and $P_{\{m\p,m\p+1\}}$. Using the trick described in
Section \ref{sectdefmf}, the result of this composition is the B-type
defect represented by the matrix factorisation associated to the 
$R=\CC[X,Z]/(X^d-Z^d)$-module
\beq
M:=\CC[X,Y,Z]/((X-\eta^mY)(X-\eta^{m+1}Y),(Y-\eta^{m\p}Z)(Y-\eta^{m\p+1}Z))\,.
\eeq
In $M$ we have the following relations:
\beqn\label{basicrels}
Y^{2+i}-\alpha XY^{1+i}-\beta X^2Y^i&=&0\\
Y^{2+i}-\alpha\p ZY^{1+i}-\beta\p Z^2Y^i&=&0\,,\nonumber
\eeqn
where we abbreviated
\beqn\label{abbr}
&&\alpha:=\eta^{-m}+\eta^{-m-1}\,,\quad \beta:=-\eta^{-2m-1}\,,\\
&&\alpha\p:=\eta^{m\p}+\eta^{m\p+1}\,,\quad\beta\p:=-\eta^{2m\p+1}\,.\nonumber
\eeqn
From this it follows in particular that the submodules of $M$ built on $Y^i$
for $i\geq 2$ are in fact submodules of those built on $1$ and
$Y$, so the task is to understand the latter, \ie the
relations in them. To start note that from \eq{basicrels} it follows
that
\beqn\label{specialrel}
0&=&\left(\alpha X-\alpha\p Z\right)Y+\left(\beta X^2-\beta\p Z^2\right)\\
&=&\left(\alpha X-\alpha\p
Z\right)\underbrace{\left(Y+{\frac{\beta}{\alpha^2}}
\left(\alpha X+\alpha\p Z\right)\right)}_{=:e_1}=0\,,\nonumber
\eeqn
where use was made of \eq{abbr}. In fact, there are no further
relations in the submodule built on $e_1$, so that the latter is just
given by ($\alpha\p/\alpha=\eta^{m+m\p+1}$)
\beq
R/(X-\eta^{m+m\p+1}Z)\,.
\eeq
To determine the remaining part of $M$, we note that \eq{basicrels}
also gives rise to \eq{specialrel} multiplied by $Y$. Substituting the
first of the equations \eq{basicrels} into the latter, one obtains,
using in particular \eq{specialrel} and \eq{abbr}
\beq
(X-\eta^{m+m\p}Z)(X-\eta^{m+m\p+1}Z)(X-\eta^{m+m\p+2}Z)=0\,,
\eeq
which is the only relation in the submodule built on $e_0:=1\in
M$. Therefore
\beq
M\cong R/(X-\eta^{m+m\p+1}Z)\oplus
R/(X-\eta^{m+m\p}Z)(X-\eta^{m+m\p+1}Z)(X-\eta^{m+m\p+2}Z)\,,
\eeq
and the defect obtained by composing the defects corresponding to the
matrix factorisations $P_{\{m,m+1\}}$ and $P_{\{m\p,m\p+1\}}$ is
represented by the sum 
\beq\label{twofactors}
P_{\{m,m+1\}}*P_{\{m\p,m\p+1\}}=P_{\{m+m\p+1\}}\oplus
P_{\{m+m\p,m+m\p+1,m+m\p+2\}}\,.
\eeq
In case $d=3$ the second summand is trivial, if $d>3$, the composition
of the two $P_I$ with $|I|=2$ generates a $P_I$ with $|I|=3$. 

Let us now consider the more general case, namely the composition of
$P_{\{m,m+1\}}$ and $P_{\{m\p,\ldots,m\p+a\}}$. The result of this
composition is the matrix factorisation associated to the $R$-module
\beq
M=\CC[X,Y,Z]/\left((X-\eta^mY)(X-\eta^{m+1}Y),\prod_{i=0}^a(Y-\eta^{m\p+i}Z)\right)\,.
\eeq
As in the special case discussed above, because of the quadratic
relation 
\beq\label{qrel}
(X-\eta^mY)(X-\eta^{m+1}Y)=Y^2-(\eta^{-m}+\eta^{-m-1})XY+\eta^{-2m-1}X^2=0\,,
\eeq
we only have to consider the submodules built on $1$ and $Y$. To
obtain the relations in them, by means of \eq{qrel} we eliminate all
$Y^i$ with $i>1$ from 
\beq
F(Y,Z)=\prod_{i=0}^a(Y-\eta^{m\p+i}Z)=0
\eeq
to obtain a relation of the form
\beq\label{rel1}
YP(X,Z)+Q(X,Z)=0\,.
\eeq
Multiplying it by $Y$ and again using \eq{qrel} gives rise to another
relation
\beqn\label{rel2}
0&=&Y^2P(X,Z)+YQ(X,Z)\\
&=&Y\left((\eta^{-m}+\eta^{-m-1})X
  P(X,Z)+Q(X,Z)\right)-\eta^{-2m-1}X^2P(X,Z)\,.\nonumber
\eeqn
Again multiplying by $Y$ one obtains a linear combination of
\eq{rel1} and \eq{rel2}, thus these two are the only relations on the
submodule built on $1$ and $Y$. 

Now, by construction
\beq
F(Y,Z=\eta^{-m\p-m-i}X)\sim\prod_{j=-a}^0(X-\eta^{m+i-j}Y)
\eeq
contains \eq{qrel} as a factor iff $1\leq i\leq a$, from which it
follows that $P(X,Z)$ and $Q(X,Z)$ have roots
$(X-\eta^{m\p+m+i}Z)$ for $1\leq i\leq a$. Since $P$ and $Q$ have
degree $a$ and $a+1$ respectively, it follows that 
\beq
P(X,Z)\sim\prod_{i=1}^a(X-\eta^{m\p+m+i}Z)\,,\quad{\rm and}\quad
Q(X,Z)=P(X,Z)q(X,Z)\,,
\eeq
where $q$ is a polynomial of degree $1$. Therefore relation \eq{rel1}
can be written as
\beq
P(X,Z)(Y+q(X,Z))=0\,,
\eeq
and using this, relation \eq{rel2} becomes
\beq
P(X,Z)\underbrace{\left(-\eta^{-2m-1}X^2-(\eta^{-m}+\eta^{-m-1})Xq(X,Z)-q^2(X,Z)\right)}_{=:S(X,Z)}=0\,.
\eeq
It remains to determine the quadratic polynomial $S(X,Z)$. For this
we note that the polynomials $F(Y,Z=\eta^{-m\p-m}X)$ and $F(Y,Z=\eta^{-m\p-m-a-1}X)$
contain factors $(X-\eta^mY)$ and $(X-\eta^{m+1}Y)$ respectively. 
In particular
\beqn
0&=&F(Y,Z=\eta^{-m\p-m}X)(X-\eta^{m+1}Y)\\
0&=&F(Y,Z=\eta^{-m\p-m-a-1}X)(X-\eta^{m}Y)\,.\nonumber
\eeqn
Making once again use of the fact that $F=YP+Q$ and the quadratic
relation \eq{qrel}, one obtains the equations
\beqn
0&=&\eta^{-m}XP(X,Z=\eta^{-m\p-m}X)+Q(X,Z=\eta^{-m\p-m}X)\\
0&=&\eta^{-m-1}XP(X,Z=\eta^{-m\p-m-a-1}X)+Q(X,Z=\eta^{-m\p-m-a-1}X)\,,
\eeqn
which can be used to determine the linear polynomial $q(X,Z)=\mu X+\nu
Z$. Namely
\beq
\mu=-\eta^ {-m}{1-\eta^a\over 1-\eta^{a+1}}\,,\quad
\nu=-\eta^{m\p+a}{1-\eta\over 1-\eta^{a+1}}\,.
\eeq
Substituting $q$ in the equation for $S$, one obtains
\beq
S(X,Z)\sim(X-\eta^{m+m\p}Z)(X-\eta^{m+m\p+a+1}Z)\,,
\eeq
and hence
\beq
M\cong\CC[X,Z]/\left(\prod_{i=1}^a(X-\eta^{m+m\p+i})\right)\oplus
\CC[X,Z]/\left(\prod_{i=0}^{a+1}(X-\eta^{m+m\p+i})\right)\,.
\eeq
Therefore, 
\beq\label{permcomplaw}
P_{\{m,m+1\}}*P_{\{m\p,\ldots,m\p+a\}}=P_{\{m+m\p+1,\ldots,m+m\p+a\}}\oplus
P_{\{m+m\p,\ldots,m+m\p+a+1\}}\,.
\eeq
If $d=a-1$ the second summand is trivial, otherwise the action of
$P_{\{m,m+1\}}$ on a $P_I$ with $|I|=r$ generates a $P_I$ with
$|I|=r+1$, and we see that by composing $P_I$ with $|I|=2$, we can
indeed generate all $P_I$. Therefore, by means of associativity,
\eq{permcomplaw} indeed determines the composition of arbitrary
permutation like defects.
Using the fusion rules ${\mathcal N}$ of
$\widehat{\mathfrak{su}}(2)_{d-2}$ one obtains:
\beq\label{gentopcomp}
P_{\{m_1,\ldots,m_1+l_1\}}*P_{\{m_2,\ldots,m_2+l_2\}}=\bigoplus_{l}{\mathcal
N}_{l_1l_2}^{l} P_{\{{1\over 2}(l_1+l_2-l)+m_1+m_2,\ldots,{1\over
2}(l_1+l_2+l)+m_1+m_2\}}\,.
\eeq
\subsubsection{Action of permutation type defects on boundary conditions}
Using the results from the previous subsection, also the action of
permutation type defects on boundary conditions is completely
determined by the action of the defects corresponding to matrix
factorisations $P_{\{m,m+1\}}$. So let us investigate the action of
such defects on boundary condition represented by 
matrix factorisations 
\beq
T_a:\quad \CC[Y]\overset{Y^a}{\underset{Y^{d-a}}{\rightleftarrows}}
\CC[Y]\,.
\eeq
The resulting boundary condition is
given by the matrix factorisation defined by the $R=\CC[X]/(W(X))$-module
\beq
M:=\CC[X,Y]/((X-\eta^mY)(X-\eta^{m+1}Y),Y^a)\,.
\eeq
The relations on this module are 
\beq\label{relaction}
Y^{2+i}-\underbrace{(\eta^{-m}+\eta^{-m-1})}_{=:\alpha}XY^{1+i}+\underbrace{\eta^{-2m-1}}_{=:-\beta}X^2Y^i=0\,,\quad
Y^a=0\,, 
\eeq
and in particular
the submodules built on $Y^i$ for $i>1$ are
submodules of the ones built on $1$ and $Y$. Therefore we only have to
determine the relations on these two submodules. 
From \eq{relaction} we obtain
\beqn
&& X^2 Y^{a-1}=0\\
&& X\left(Y^{a-1}+{\beta\over\alpha}XY^{a-2}\right)=0\,,\nonumber
\eeqn
and inductively:
\beqn
&& X^{i+2}Y^{a-i-1}=0\\
&& X^i\left(Y^{a-i}- {\sum_{j=0}^{i-1}\eta^{j-m}\over\sum_{j=0}^{i}\eta^j}XY^{a-i-1}\right)=0\,.\nonumber
\eeqn
In particular:
\beqn
&& X^{a+1}1=0\\
&& X^{a-1}\left(Y-
  {\sum_{j=0}^{a-2}\eta^{j-m}\over\sum_{j=0}^{a-1}\eta^j}X\right)=0\nonumber
\eeqn
and therefore, as an $R$-module
\beq
M\cong R/X^{a-1}R\oplus R/X^{a+1}R\,.
\eeq
Thus, applying the defect corresponding to the matrix factorisation
$P_{\{m,m+1\}}$ to the boundary condition associated to $T_a$
($0<a<d$) results
in the boundary condition described by the matrix factorisation
\beq
P_{\{m,m+1\}}*T_a=T_{a-1}\oplus T_{a+1}\,.
\eeq
If $a=1$ or $a=d-1$ one of the summands is a trivial matrix factorisation. 
Using \eq{gentopcomp}, one can obtain the action of arbitrary $P_I$ to
be
\beq\label{gentopact}
P_{\{m,\ldots,m+l\}}*T_a=\bigoplus_{b}{\mathcal N}_{la}^b T_b\,.
\eeq

\subsubsection{Action of tensor product type defects}
In this subsection we will discuss the action of tensor product (TP)
type defects on
boundary conditions and other TP type defects. Let us start with the
discussion of the application of the defect corresponding to $T_{a,b}$
on the boundary condition represented by $T_\beta$. Indeed,
\beq\label{eqtpaction}
T_{a,b}*T_\beta=\left(T_a(X)\otimes\bar T_b(Y)\right)\otimes T_\beta(Y)\,.
\eeq
Now, let us assume that the minimum $m=\min(b,\beta)$ satisfies $m\leq
d-m$. This can always be achieved by shifting both $T_b\mapsto T_b[1]$ and
$T_\beta\mapsto T_\beta[1]$, which does not affect \eq{eqtpaction}. Consider
the case $\beta=m$. The matrix factorisation \eq{eqtpaction} is
isomorphic (up to trivial matrix factorisations) to the one arising
from the resolution of the module
\beqn\label{TPM}
M&=&\coker(t^{a,b}_1\otimes\id,-\id\otimes t_1^\beta)\\
&=&\coker\left(\begin{array}{cccc}x^a&y^b&-y^\beta&0\\y^{d-b}&x^{d-a}&0&-y^\beta\end{array}\right)\nonumber\\
&=&\coker\left(\begin{array}{cccc}x^a&0&-y^\beta&0\\0&x^{d-a}&0&-y^\beta\end{array}\right)\nonumber\\
&\cong&\CC[X,Y]/(X^a,Y^\beta)\oplus\CC[X,Y]/(X^{d-a},Y^\beta)\nonumber
\,,
\eeqn
where it was used that $\beta\leq b,d-b$. Therefore for $\beta<d-\beta,b,d-b$
\beq
T_{a,b}*T_\beta=\left(T^a\oplus T^{d-a}\right)^{\oplus\beta}
=\left(T^a\oplus T^{a}[1]\right)^{\oplus\beta}
\,.
\eeq
If $b<\beta$ one can use the associativity of the tensor product of
matrix factorisations to obtain a module $M$ with cokernel
representation as in \eq{TPM} with $\beta$ and $b$ interchanged. (Also
some irrelevant signs are different because one of the $T_b,T_\beta$
appearing in \eq{eqtpaction} has a bar.) Thus, in the same way, one
arrives at the result for arbitrary $b$ and $\beta$:
\beq
T_{a,b}*T_\beta=\left(T^a\oplus T^a[1]\right)^{\oplus\min(b,\beta,d-b,d-\beta)}\,.
\eeq
Analogously one can deal with the composition of TP like defects
to obtain
\beq
T_{a,b}*T_{\beta,\gamma}=\left(T_{a,\gamma}\oplus
  T_{a,\gamma}[1]\right)^{\oplus\min(b,\beta,d-b,d-\beta)}\,.
\eeq
Since $T_{\beta,\gamma}=T_\beta(Y)\otimes\bar T_\gamma(Z)$ is a tensor
product matrix factorisation, this result indeed can be easily
obtained from the action of $T_{a,b}$ on boundary conditions,
namely
\beq
T_{a,b}*T_{\beta,\gamma}=(T_{a,b}*T_\beta)\otimes \bar T_\gamma\,,
\eeq
and this trick can in fact also be used to deduce the action of
permutation type defects on TP like defects from their action on
boundary conditions:
\beq
P_{I}*T_{\beta,\gamma}=(P_{I}*T_\beta)\otimes \bar T_\gamma\,.
\eeq
\subsubsection{$W=X^d+Z^2$}\label{sectlgz2}
Here we would like to extend the previous analysis to Landau-Ginzburg models
with superpotentials $W=X^d+Z^2$. Defects between these models correspond to matrix
factorisations of $X^d+Z^2-Y^d-U^2$. The obvious generalisations of
the factorisations \eq{defpfact} are just tensor products of the
factorisations $P_I^d(X,Y)$ of $X^d-Y^d$ and factorisations 
$P_J^2(Z,U)$ of $Z^2-U^2$. Obviously $J$ can be chosen to
consist either of $0$ or $1$, and these factorisations are symmetry
defects with respect to the $\ZZ_2$ generated by changing the sign of
the respective superfield. 
We denote the tensor products as
\beq\label{defmfz2}
P_{I}^\pm:=P_I^d(X,Y)\otimes P_{\{\pm 1 -1\}}^2(Z,U)\,.
\eeq
Note that not all of these factorisations are independent. 
Since the tensor product of two shifted matrix
factorisations is equivalent to the tensor product of the unshifted
ones, $P[1]\otimes Q[1]\cong P\otimes Q$, we have
\beq
P_{I}^\pm\cong P_{\{0,\ldots d-1\}-I}^\mp\,,
\eeq
and all these defects can be expressed in terms of $P_I^+$ only. This is
expected from Kn\"orrer periodicity, which states that the category of
matrix factorisations of a polynomial and that one of the same polynomial to
which two squares are added are equivalent. In particular, the
structure of defects
in theories with superpotential $W=X^d$ and $W=X^d+Z^2$ should
coincide\footnote{Of course one can also study defects between Landau-Ginzburg
models with superpotentials with $W=X^d$ and $W=X^d+Z^2$, which would
then correspond to matrix factorisations of $X^d+Z^2-Y^d$. The
structure of these kinds of defects is different from the ones between
models of the same type, but we will refrain from discussing them
here.}.

Because of the tensor product structure, composition of these defects
can  easily be reduced to the one of the tensor factors. Thus from
\eq{gentopcomp} and the obvious $\ZZ_2$-composition of the symmetry
defects, one deduces
\beq\label{gentopcompz2}
P_{\{m_1,\ldots,m_1+l_1\}}^\sigma*P_{\{m_2,\ldots,m_2+l_2\}}^\rho=\bigoplus_{l}{\mathcal
N}_{l_1l_2}^{l} P_{\{{1\over 2}(l_1+l_2-l)+m_1+m_2,\ldots,{1\over
2}(l_1+l_2+l)+m_1+m_2\}}^{\sigma\rho}\,.
\eeq
We would like to study how defects corresponding to these matrix
factorisations act on boundary conditions. Corresponding to matrix
factorisations of $X^d+Z^2$, the latter come in two classes \cite{Kapustin:2003rc}.
Firstly, there are the obvious tensor product matrix factorisations
\beq
\Theta_a:=T_a^d(X)\otimes T_1^2(Z)\,,
\eeq
where as before
\beq
T_a^d(X):\quad \CC[X]\overset{X^a}{\underset{X^{d-a}}{\rightleftarrows}}
\CC[X]\,.
\eeq
These factorisations are not ``oriented'' in the sense that
$\Theta_a\cong\Theta_a[1]\cong\Theta_{d-a}$. The action of the defects
$P_I^\pm$ on them can again be decomposed into the action of the
respective tensor factors and with \eq{gentopact} one obtains
\beq\label{gentopactz2}
P_{\{m,\ldots,m+l\}}^\pm*\Theta_a=\bigoplus_{b}{\mathcal N}_{la}^b
\Theta_b\,.
\eeq
For even $d$ however the factorisations $\Theta_{d\over 2}$ are
reducible. They split up
\beq
\Theta_{d\over 2}\cong \Psi^+\oplus \Psi^-
\eeq
into the two additional rank-one factorisations
\beq
\Psi^\pm:\quad \CC[X,Z]\overset{\psi_1^\pm}{\underset{\psi_0^\pm}{\rightleftarrows}}
\CC[X,Z]\,,\qquad
\psi_1^\pm=\left(X^{d\over 2}\mp iZ\right)\,,\quad \psi_0^\pm=\left(X^{d\over 2}\pm iZ\right)\,.
\eeq
In contrast to the $\Theta_i$, the $\Psi^\pm$ are oriented; they satisfy
$\Psi^\pm[1]\cong\Psi^\mp\ncong\Psi^\pm$, and the action of the
defects associated to the $P_I^\pm$ on the corresponding  
boundary conditions is more complicated. $P_{\{m\}}^\pm$ for instance
is a symmetry defect and as discussed in Section \ref{sectgrouplike}
acts on any matrix factorisation by replacing 
\beq
X\mapsto\eta^{-m}X\,,\quad Z\mapsto \pm Z\,.
\eeq
In particular, 
\beq\label{symactpsi}
P^\sigma_{\{m\}}*\Psi^\rho=\Psi^{\sigma\rho\eta^{md\over 2}}
=\left\{\begin{array}{ll}\Psi^{\sigma\rho}\,,& m\;{\rm even}\\
\Psi^{-\sigma\rho}\,,&m\;{\rm odd}\end{array}\right.\,.
\eeq
In view of the fact that also the $P^\pm_I$ with $|I|>2$ are
generated by the composition of those with $|I|\leq 2$ (\cf
\eq{gentopcompz2}), we again only have to analyse the action of the 
$P^\pm_{\{m,m+1\}}$ on $\Psi^\pm$ by hand. To do this, we note that the
result of $P^\sigma_{\{m,m+1\}}*\Psi^\rho$ is the matrix factorisation
obtained from $R=\CC[X,Z]/(X^d+Z^2)$-free resolutions of the
module
\beqn
M&=&\CC[X,Y,Z,U]/\left((Y^{d\over 2}-i\rho U),(Z-\sigma
U),(X-\eta^mY)(X-\eta^{m+1}Y)\right)\\
&\cong&\CC[X,Y,Z]/\left((Y^{d\over 2}-i\sigma\rho Z),(X-\eta^mY)(X-\eta^{m+1}Y)\right)\,.
\eeqn
Because of the quadratic relation $(X-\eta^mY)(X-\eta^{m+1}Y)=0$ in
$M$, $i$th powers of $Y$ with $i\geq 2$ can be expressed as
\beq\label{PQ1}
Y^i=P_i(X)+Y Q_i(X)\,.
\eeq
Inductively one easily finds that
\beqn\label{PQ2}
&&P_i(X)=p_i X^i\,,\quad p_i=-\eta^{-(m+1)i+1}(1+\eta+\ldots+\eta^{i-2})\,,\\
&&Q_i(X)=q_i X^{i-1}\,,\quad q_i=\eta^{-(m+1)(i-1)}(1+\eta+\ldots+\eta^{i-1})\,.\nonumber
\eeqn
Therefore, $M$ collapses to a submodule of
$\CC[X,Z]\oplus Y\CC[X,Z]$, and the only task is to find the relations
in it. These come from the relations
\beqn
&&Y^{{d\over 2}-2}X^2-\eta^m(1+\eta)Y^{{d\over
2}-1}X+i\eta^{2m+1}\sigma\rho Z=0\,,\label{redrel1}\\
&&Y^{{d\over 2}-1}X^2-i\eta^m(1+\eta)\sigma\rho
XZ+i\eta^{2m+1}\sigma\rho YZ=0\,,\label{redrel2}
\eeqn
which are obtained by substituting $Y^{d\over 2}=i\sigma\rho Z$ into
 $Y^i(X-\eta^m Y)(X-\eta^m Y)=0$ for $i={d\over 2}-2$ and
$i={d\over 2}-1$ respectively.
Using \eq{PQ1}, \eq{PQ2} and the explicit form of the $p_i$ and $q_i$
these equations can be written as
\beqn
&&\left(q_{{d\over 2}-1}X^{d\over 2}+i\eta^{2m+1}\sigma\rho Z\right)
+Y\left(-\eta^{2m+1}q_{d\over 2} X^{{d\over 2}-1}\right)=0\,,\\
&&\left(p_{{d\over 2}-1}X^{{d\over 2}+1} -i\eta^m(1+\eta)\sigma\rho
XZ\right)
+Y\left(q_{{d\over 2}-1} X^{d\over 2}+i\eta^{2m+1}\sigma\rho
Z\right)=0\,.
\eeqn
Regarding $\CC[X,Z]\oplus Y\CC[X,Z]$ as $\CC[X,Z]^2$, $M$ is
isomorphic to the cokernel of the matrix
\beq
O=\left(\begin{array}{cc}
q_{{d\over 2}-1}X^{d\over 2}+i\eta^{2m+1}\sigma\rho Z\phantom{**} &
p_{{d\over 2}-1}X^{{d\over 2}+1} -i\eta^m(1+\eta)\sigma\rho XZ \\
-\eta^{2m+1}q_{d\over 2} X^{{d\over 2}-1} &
q_{{d\over 2}-1} X^{d\over 2}+i\eta^{2m+1}\sigma\rho
Z\end{array}\right)\,.
\eeq
By means of elementary row and column transformations this matrix can
be brought into the form
\beq
\left(\begin{array}{cc}
i\sigma\rho Z &
X^{{d\over 2}+1}\left(p_{{d\over 2}-1}+{q_{{d\over
2}-1}^2\over\eta^{2m+1}q_{d\over 2}}\right)+i\sigma\rho
XZ\left(-\eta^m(1+\eta)+2{q_{{d\over 2}-1}\over q_{d\over 2}}\right)\\
-q_{d\over 2} X^{{d\over 2}-1} &
i\eta^{2m+1}\sigma\rho Z
\end{array}\right)
\eeq
Using the explicit formulas for the $q_i$ and $p_i$, in particular
$q_{d\over 2}^{-1}={1\over 2}(1-\eta)\eta^{(m+1)({d\over 2}-1)}$, one
can show that the upper right entry of this matrix indeed simplifies
to
\beq
{-\eta^{2m+1}\over q_{d\over 2}}X^{{d\over 2}+1}\,,
\eeq
and again using elementary row and column transformations $O$ can be
brought into the form
\beq
O\mapsto
\left(\begin{array}{cc}
X^{{d\over 2}-1} & -Z\\
Z & X^{{d\over 2}+1} 
\end{array}\right)\,,
\eeq
which is easily recognised as the matrix $\theta_{{d\over 2}-1}^1$ of
the matrix factorisation $\Theta_{{d\over 2}-1}$. Thus, 
\beq
M\cong\coker(\theta_{{d\over 2}-1}^1)\,,
\eeq
and
\beq\label{defso}
P_{\{m,m+1\}}^\sigma*\Psi^\rho=\Theta_{{d\over 2}-1} = \frac{1}{2}
\sum_l {\mathcal N}_{1 \frac{d}{2}}^l \Theta_l \,.
\eeq
By means of the composition \eq{gentopcompz2} this determines the
action of all $P_I^\pm$ on the $\Psi^\pm$. For $l_1$ odd, it is 
straightforward to derive
\beq\label{deflodd}
P^\sigma_{\{m,\dots, m+l_1\}} *\Psi^\rho = 
\frac{1}{2} \sum_l {\mathcal N}_{l_1 \frac{d}{2}}^l \Theta_l \, .
\eeq
The simplest case for $l_1$ even is obviously $l_1=0$, which has been treated
above, \cf \eq{symactpsi}. The next simple case is $l_1=2$, for which the action of the defect
can be obtained from
\beq
P^\sigma_{\{ m,m+1 \} } * P^{\sigma'}_{\{ m',m'+1 \} } \Psi^\rho=
P^\sigma_{\{ m,m+1 \} }*\Theta_{{d\over 2}-1}=\Psi^+\oplus\Psi^-\oplus\Theta_{{d\over
2}-2}\,.
\eeq
Here, we used that the factorisation $\Theta_{\frac{d}{2}}$ is reducible
and can be decomposed into $\Psi^+$ and $\Psi^-$. Applying \eq{gentopcompz2}
we obtain
\beq
P^{\sigma\sigma'}_{\{ m+m',m+m'+1,m+m'+2 \}} * \Psi^\rho
=\Psi^{(-1)^{m+m\p}\sigma \sigma' \rho} \oplus \Theta_{\frac{d}{2}-2}
\eeq
This immediately generalises to
\beq\label{defleven}
P^\sigma_{\{ m, \dots, m+l_1 \}} * \Psi^\rho =
\Psi^{(-1)^{m}\sigma \rho} \oplus \frac{1}{2} \sum_l
{\mathcal N}_{\frac{d}{2} l_1}^{l} \Theta_{l} \quad {\rm for} \ l_1 \ {\rm even} \, .
\eeq
\subsection{CFT approach}\label{CFTapproach}
In the IR, the Landau-Ginzburg model with one chiral superfield
and superpotential $W(X)=X^d$ and the one with an additional
superfield and superpotential $W(X,Z)=X^d+Z^2$ both flow to versions of 
the unitary superconformal minimal
model ${\mathcal M}_k$, $k=d-2$ with A-type modular invariant
partition function. The two versions only differ in the definition of 
$(-1)^F$ on the Ramond-sectors.

The conformal field theories ${\mathcal M}_k$ are rational with respect to the $N=2$ super
Virasoro algebra at central charge $c_k={3k\over k+2}$. In fact, the
bosonic part of this algebra can be realised as the coset W-algebra
\beq\label{coset}
\left({\rm SVir}_{c_k}\right)_{\rm bos} = 
{\widehat{\mathfrak{su}}(2)_k\oplus\widehat{\mathfrak{u}}(1)_4\over \widehat{\mathfrak{u}}(1)_{2k+4}}\,,
\eeq  
and the respective coset CFT can be obtained from ${\mathcal M}_k$ by
a non-chiral GSO projection. 

The Hilbert space $\HH^k$ of ${\mathcal M}_k$ decomposes into irreducible
highest weight representations of holomorphic and antiholomorphic
super Virasoro algebras, but it is convenient to decompose it further
into irreducible highest weight representations $\VV_{[l,m,s]}$ of the
bosonic subalgebra \eq{coset}. These representations are labelled by 
\beq
[l,m,s]\in\mathcal{I}_k:=\{(l,m,s)\,|\,0\leq l\leq
k,\,m\in\ZZ_{2k+4},\,s\in\ZZ_4,\,l+m+s\in 2\ZZ\}/\sim\,,
\eeq
where $[l,m,s]\sim[k-l,m+k+2,s+2]$ is the field identification. The highest weight representations
of the full super Virasoro algebra are given by
\beq
\VV_{[l,m]}:=\VV_{[l,m,(l+m)\,{\rm
    mod}\,2]}\oplus\VV_{[l,m,(l+m)\,{\rm mod}\,2+2]}\,.
\eeq
For $(l+m)$ even $\VV_{[l,m]}$ is in the NS-, for $(l+m)$ odd in the R-sector.
Here $[l,m]\in{\mathcal J}_k:=\{(l,m)\,|\,0\leq l\leq
k,\,m\in\ZZ_{2k+4}\}/\sim$, $[l,m]\sim[k-l,m+k+2]$. 
The Hilbert spaces of ${\mathcal
M}_k$ in the NSNS- and RR-sectors then read
\beq
\HH^k_{NSNS} 
\cong\bigoplus_{\stackrel{[l,m]\in{\mathcal J}_k}{l+m\,{\rm
      even}}}\VV_{[l,m]}\otimes\overline{\VV}_{[l,m]} \,,\qquad
\HH^k_{RR}
\cong\bigoplus_{\stackrel{[l,m]\in{\mathcal J}_k}{l+m\,{\rm
      odd}}}\VV_{[l,m]}\otimes\overline{\VV}_{[l,m]} \,.
\eeq
In this section we would like to discuss topological defects
in the supersymmetric model
${\mathcal M}_k$ preserving B-type supersymmetry. Located on the real line $z=z^*$
they impose the following gluing conditions
\beq\label{gluingcond}
\left.
\begin{array}{r}
T(z)-T(z^*)\\
\overline T(\bar z)-\overline T(\bar z^*)\\
G^{\pm}(z)-\eta G^{\pm}(z^*)\\
\overline G^{\pm}(\bar z)-\bar\eta\overline G^{\pm}(\bar z^*)
\end{array}
\right\}\rightarrow 0\quad{\rm for}\quad z-z^*\rightarrow 0\,,
\eeq
for $\eta,\bar\eta\in\{\pm 1\}$. 
Representing the defects as operators ${\mathcal D}:\HH^k\rightarrow\HH^k$
the gluing conditions lead to commutation relations
\beqn\label{commreldef}
\left[L_n,{\mathcal D}\right]=&0&=\left[\overline{L}_n,{\mathcal D}\right]\\
G^{\pm}_r{\mathcal D}-\eta{\mathcal D}G^{\pm}_r=&0&=
\overline{G}^{\pm}_r{\mathcal D}-\bar\eta{\mathcal
  D}\overline{G}^{\pm}_r\,,\nonumber
\eeqn
for all $n\in\ZZ$ and all $r\in\ZZ+{1\over 2}$ ($r\in\ZZ$) in the NS-
(R-) sectors. We will furthermore require
${\mathcal D}$ to commute with $(-1)^F$, which in general might be
defined differently on both sides of the defect. In this paper, the discussion
will be
restricted  to the case in which the action of $(-1)^F$ on both sides of
the defect is the same, \ie we will only discuss defects between the
same type of models. 
Composing ${\mathcal D}$ with $(-1)^F$ results in
an operator satisfying gluing conditions with opposite $\eta$ and $\bar\eta$.
Likewise, $\eta$ and $\bar\eta$ can be changed separately if 
$(-1)^{F_L}$ and $(-1)^{F_R}$ are on their own symmetries of the 
theory\footnote{In fact, the operators $(-1)^F$, $(-1)^{F_L}$ and
$(-1)^{F_R}$ are indeed associated to topological defects as well, and
composition with ${\mathcal D}$ can be interpreted as fusion of the respective
defects.}.

Since ${\mathcal M}_k$ is a diagonal RCFT with respect to the $N=2$
algebra,
standard techniques can be used to construct the defect operators.
First, Schur's lemma implies that for $\eta=\bar\eta$
${\mathcal D}$ has to be a linear
combination
\beq\label{defoptsuper}
{\mathcal D}=\sum_{[l,m]} {\mathcal D}^{[l,m]} {\rm P}_{[l,m]}
= {\mathcal D}_{NSNS} + {\mathcal D}_{RR} \ .
\eeq
of 
projectors ${\rm P}_{[l,m]}$ on the irreducible
representations $\VV_{[l,m]}\otimes\overline{\VV}_{[l,m]}$.
From this it is simple to obtain defects corresponding to other choices
of $\eta$ by  composing with $(-1)^{F_L}$ or $(-1)^{F_R}$.
Note that this formula combines the action of the defect on both NSNS-
($l+m$ even) as well as RR-sectors ($l+m$ odd). 
At this point we assume that we are dealing with defects
between the same type of model, \ie with the same definition of
$(-1)^F$. Namely, in contrast to defects between the same version of minimal
models, defects between the two different versions are linear
  combinations of intertwiners between representations
  $\VV_{[l,m,s]}\otimes\overline{\VV}_{[l,m,\bar s]}$ and 
  $\VV_{[l,m,s]}\otimes\overline{\VV}_{[l,m,-\bar s]}$.

Indeed, also for the case of defects between the same version of
minimal models it is useful to write the defect operators as sums over projectors
${\rm P}_{[l,m,s,\bar s]}$ of the modules
$\VV_{[l,m,s]}\otimes\overline{\VV}_{[l,m,\bar s]}$ of the bosonic subalgebra 
\beq
{\mathcal D}=\sum_{\stackrel{[l,m,s],\bar s}{s-\bar s\,{\rm
      even}}}{\mathcal D}^{[l,m,s,\bar s]} {\rm P}_{[l,m,s,\bar s]}\,,
\eeq
where it is understood that 
\beq
{\mathcal
  D}^{[l,m,s+2,\bar s]}=\eta{\mathcal D}^{[l,m,s,\bar
  s]}\quad{\rm and}\quad 
{\mathcal
  D}^{[l,m,s,\bar s+2]}=\bar \eta{\mathcal D}^{[l,m,s,\bar
  s]}\,.
\eeq
The possible linear combinations of projectors are restricted
by sewing relations which ensure that correlation functions do not
depend on the different ways in which surfaces can be sewn
together. In particular there is a sewing relation similar to Cardy's constraint for boundary
conditions (see \eg \cite{Petkova:2000ip}). The standard solution, which can
also be obtained via the folding trick from permutation boundary
conditions is given by
\beq\label{qdim}
{\mathcal D}_{[ L, M,
S,\bar S]}^{[l,m,s,\bar s]}=e^{-i\pi {\bar S(s+\bar s)\over 2}}{S_{[ L, M,
  S-\bar S][l,m,s]}\over
S_{[0,0,0],[l,m,s]}}\,,
\eeq
where the different defects have been labelled by $[ L,
M, S,\bar S]$ with $[L,M,S-\bar S]\in {\mathcal I}_k$, and 
\beq
S_{[L,M,S][l,m,s]}={1\over k+2}e^{-i\pi{Ss\over 2}}e^{i\pi{Mm\over
k+2}}\sin\left(\pi{(L+1)(l+1)\over k+2}\right)
\eeq
is the modular $S$-matrix for the coset representations
$\VV_{[l,m,s]}$. Obviously, the possible choices of $S$ and $\bar S$ are 
determined
by $\eta$ and $\bar\eta$ in the usual way,  $\eta=(-1)^S$ and $\bar\eta
=(-1)^{\bar{S}}$. The defect does not change
under $(S,\bar S)\mapsto(S+2,\bar S+2)$.  

Since these defects are topological we can bring them together to
obtain new defects. From \eq{commreldef} it is clear that this
operation preserves the gluing conditions so that the result 
will again be a B-type defect. The twist parameters $\eta$ and
$\bar\eta$ are multiplicative. 
On the level of defect operators this operation 
just amounts to their composition. 
Using the fact that the quantum dimensions \eq{qdim} form representations
of the respective fusion rules ${\mathcal N}$, one easily obtains the composition law
\beqn\label{CFTdefcomp}
{\mathcal D}_{[ L_1, M_1, S_1,\bar S_1]}{\mathcal D}_{[ L_2,
M_2, S_2,\bar S_2]}&=&\!\!\!\sum_{[ L, M, S-\bar S]\in{\mathcal I}_k,\bar S}\!\!\!{\mathcal
N}_{[ L_1, M_1, S_1-\bar S_1][ L_2, M_2, S_2-\bar S_2]}^{[
L, M, S-\bar S]}\delta^{(4)}_{\bar S_1+\bar S_2,\bar S}{\mathcal D}_{[ L, M,
S,\bar S]}\nonumber\\
&=&\sum_{ L}{\mathcal N}_{ L_1  L_2}^{ L}{\mathcal D}_{[ L,
M_1+ M_2, S_1+ S_2,\bar S_1+\bar S_2]}\,.
\eeqn
Note that for $L=0$ these defects are group-like. The
defect labels $[0,M,S,\bar S]$ correspond to simple currents, and
their fusion determines the composition of the corresponding defects.
\subsubsection{Action on boundary conditions}
Next, we would like to discuss the action of these topological B-type
defects on B-type boundary conditions. On the real
line the latter impose gluing conditions
\beq\label{gluingcondbc}
\left.
\begin{array}{r}
T(z)-\overline T(\bar z)\\
G^{\pm}(z) -\eta \overline G^{\pm}(\bar z)
\end{array}
\right\}\rightarrow 0\quad{\rm for}\quad z-z^*\rightarrow 0\,,
\eeq
translating into the relations 
\beqn\label{commrelbs}
&&(L_n-\overline L_{-n})\kket{B}=0\,.\\
&&(G^\pm_r  -i\eta  \overline G^\pm_{-r})\kket{B}=0\nonumber
\eeqn
for the respective boundary states $\kket{B}$. The choice of sign
$\eta\in\{\pm 1\}$ 
in the gluing conditions for the supercurrents corresponds to the
choice of different spin structures. Modules
$\VV_{[l,m,s]}\otimes\overline{\VV}_{[l,m,s]}$ support Ishibashi
states $\iket{[l,m,s]}_B$ solving the gluing conditions \eq{commrelbs} if
$[l,m,s]\sim[l,-m,-\bar s]$. Thus, there are Ishibashi states
$\iket{[l,0,s]}_B$ for all $[l,0,s]\in{\mathcal I}_k$. In
case $k$ is even 
there are additional Ishibashi states $\iket{[{k\over 2},{k+2\over
    2},1]}_B$.

Apart from the gluing conditions \eq{commrelbs} above, the boundary condition should
also preserve $\ZZ_2$-fermion number, which means that 
$(-1)^{F}\kket{B}=\kket{B}$. Since the two CFTs corresponding to the
Landau-Ginzburg models with superpotentials $W=X^d$ and $W=X^d+Z^2$
differ by the definition of $(-1)^F$, we have to treat the two cases
separately.
 
\medskip
\noindent{\bf Case 1: $W=X^d$}
\medskip

\noindent Let us start with the CFT associated to the superpotential
$W=X^d$. In this model $(-1)^F$ acts on 
$\VV_{[l,m,s]}\otimes\overline{\VV}_{[l,m,\bar s]}$ as multiplication
by $(-1)^{s+\bar s\over 2}$, and hence only Ishibashi states
$\iket{[l,0,s]}_B$ can contribute to B-type boundary
states\footnote{The relevant GSO-projection in this model is of type
  $0A$ projecting onto the subspace
  $\HH_k^{0A}\cong\bigoplus(\VV_{[l,m,s]}\otimes\overline{\VV}_{[l,m,-s]})$.}.
The standard construction yields boundary states
\beqn\label{Bbound}
\kket{[ L, M, S]}^{\rm NS}_B&=&\kket{[ L, M+2, S]}^{\rm NS}_B
=\sqrt{{{\scriptstyle 2(k+2)}}}\!\! \sum_{\stackrel{[l,0,s]\in{\mathcal I}_k}{s\;{\rm even}}}\!\!\!{S_{[
L, M, S][l,0,s]}\over\sqrt{S_{[0,0,0][l,0,s]}}}\iket{[l,0,s]}_B\phantom{**}\\
\kket{[ L, M, S]}^{\rm R}_B&=&\kket{[ L, M+2, S]}^{\rm R}_B
=\sqrt{\scriptstyle 2(k+2)}\!\! \sum_{\stackrel{[l,0,s]\in{\mathcal I}_k}{s\;{\rm odd}}}\!\!\!{S_{[
L, M, S][l,0,s]}\over\sqrt{S_{[0,0,0][l,0,s]}}}\iket{[l,0,s]}_B\nonumber\,,
\eeqn
for every $[L,M,S]\in{\mathcal I}_k$, 
where we have specified both, the NSNS- as well as the RR-components. 
The boundary states in the GSO projected theory can be obtained by
adding RR- and NSNS-part of the boundary state with a normalisation factor
$\frac{1}{\sqrt{2}}$.

A shift by $2$ in the $S$ labels inverts
the sign in front of the RR-sector Ishibashi states and hence corresponds 
to a brane-anti-brane map. 
Similarly as in the defect case, $S$ mod $2$ is
given by $\eta=(-1)^S$ in the gluing conditions 
\eq{gluingcondbc} above.

In case $k$ is odd, all boundary states are
oriented, \ie they have non-trivial RR-components and are therefore not
invariant under the brane-anti-brane map. If $k$ is even, 
the boundary states $\kket{[{k\over 2},{k\over 2}-S,S]}_B$ have
vanishing RR-component, and are therefore unoriented.

Moving the topological B-type defects constructed above to a boundary with
B-type boundary condition amounts to applying the corresponding defect operators
${\mathcal D}_{[ L_1, M_1, S_1,\bar S_1]}$ to the respective boundary state
$\kket{[ L_2, M_2, S_2]}_B$. From \eq{commreldef} and
\eq{commrelbs} it is obvious that the resulting states
again satisfy B-type gluing conditions and preserve $(-1)^F$. 
Furthermore, sewing relations
ensure that these states are again boundary states. 
Direct calculation 
yields
\beqn\label{CFTdefoperation}
{\mathcal D}_{[ L_1, M_1, S_1,\bar S_1]}\kket{[ L_2, M_2, S_2]}_B
&=&\!\!\!\sum_{[ L, M, S]\in{\mathcal I}_k}\!\!\!{\mathcal
N}_{[ L_1, M_1, S_1-\bar S_1][ L_2,M_2, S_2]}^{[
L, M, S]}\kket{[ L, M, S]}_B\\
&=&\sum_{ L}{\mathcal N}_{ L_1 L_2}^{ L} \kket{[
L, M_1+ M_2, S_1-\bar S_1+ S_2]}_B\,.\nonumber
\eeqn
From this one immediately deduces that
defects with $S_1-\bar S_1=0$ map branes to branes, hence are
orientation preserving, whereas defects with $S_1-\bar S_1=2$ reverse brane
orientation. Defects with odd $S_1-\bar S_1$ flip the spin structure
compatible with the boundary condition. 

\medskip
\noindent{\bf Case 2: $W=X^d+Z^2$}

\noindent
In the model corresponding to $W=X^d+Z^2$, $(-1)^F$ acts on 
$\VV_{[l,m,s]}\otimes\overline{\VV}_{[l,m,\bar s]}$ as multiplication
by $(-1)^{s-\bar s\over 2}$, which only leaves the Ishibashi states
$\iket{[l,0,s]}_B$ for even $s$ and $\iket{[{k\over 2},{k+2\over
    2},\pm 1]}_B$ invariant\footnote{The corresponding GSO-projection is
  of type $0B$ and projects onto the subspace 
$\HH_k^{0B}\cong\bigoplus(\VV_{[l,m,s]}\otimes\overline{\VV}_{[l,m,s]})$.}. 
Since the models corresponding to the
superpotentials $W=X^d$ and $W=X^d+Z^2$ are $\ZZ_2$-orbifolds of each
other \cite{Kapustin:2003rc}, the boundary states of one of the models
can be obtained from the ones of the other by means of a standard
orbifold construction. Applying this construction to the boundary
states \eq{Bbound} one obtains \cite{Maldacena:2001ky}
\beqn\label{Bboundz2}
\kket{[ L, M, S]}_B^{NS}&=&\kket{[L,M+2,S]}_B^{NS}=\kket{[L,M,S+2]}_B^{NS}\\
&=&2^{1-\delta_{L,{k\over 2}}}\sqrt{\scriptstyle k+2}
\sum_{\stackrel{[l,0,s]\in{\mathcal I}_k}{s\;{\rm even}}}
{S_{[L, M, S][l,0,s]}\over\sqrt{S_{[0,0,0][l,0,s]}}}
\iket{[l,0,s]}_B\nonumber\\
\kket{[ L, M, S]}_B^R&=&\kket{[L,M+2,S]}_B^R
=\delta_{L,{k\over 2}} \sqrt{\scriptstyle 2(k+2)} e^{-\frac{i\pi S^2}{2}}
\sum_{s=\pm 1} e^{-\frac{i\pi Ss}{2}} 
\iket{[\frac{k}{2},\frac{k+2}{2},s]}_B\,.\nonumber
\eeqn
Note that for $L\neq {k\over 2}$, the $\ZZ_2$-orbifold projects out
the RR-components of the respective boundary states, so that 
$\kket{[L,M,S]}_B^R=0$ for $L\neq {k\over 2}$. Thus, the boundary
states associated to such $[L,M,S]$ 
are not oriented, and only depend on $S\;{\rm mod}\;2$, which 
distinguishing the spin structures $\eta=(-1)^S$.  
Only in case of even $k$ do there exist oriented
boundary states. These emanate from boundary states
with $L={k\over 2}$ in the unorbifolded theory 
which are invariant under the orbifold group and
therefore pick up twisted RR-sector 
contributions upon orbifolding. They are not invariant with respect to
$S\mapsto S+2$.

Since the boundary states \eq{Bboundz2} with $L\neq {k\over 2}$ 
just correspond to $\ZZ_2$-orbits of boundary states \eq{Bbound} of
the unorbifolded model, one can immediately conclude from
\eq{CFTdefoperation} that the action
of the defects on these states is given by
\beqa\label{deflongorbit}
&& {\mathcal D}_{[L_1,M_1,S_1,\bar S_1]}  \kket{[L_2,M_2,S_2]}_B
= \sum_{L\neq \frac{k}{2}} {\mathcal N}_{L_1 L_2}^L \kket{L,M_1+M_2,S_1+\bar S_1+S_2}_B \\  \nonumber
&&\quad+N_{L_1 L_2}^{\frac{k}{2}} 
\left(\kket{[\frac{k}{2},M_1+M_2,S_1+\bar{S}_1 + S_2]}_B +
\kket{[\frac{k}{2},M_1+M_2,S_1+\bar{S}_1 + S_2+2]}_B \right)
\eeqa
No RR-sector contribution can arise, and therefore only unoriented
boundary states can emerge from this operation. In particular, if
$k/2$ is contained in the fusion of $L_1$ and $L_2$ the sum of the two
short orbit boundary states appears.
Since the branes remain  unoriented,
defects whose $S_1+\bar{S}_1$ differs by $2$ 
act in the same way.

More interesting is the action on the unoriented boundary states with 
$L=\frac{k}{2}$. The action of a defect
${\mathcal D}_{[L_1,M_1,S_1,\bar S_1]}$ on the NS-component of a boundary state
$\kket{[{k\over 2},M,S]}_B$ 
is simply given by one half of (\ref{deflongorbit}) with $L_2=k/2$. Note that if
$L$ appears in the fusion of $k/2$ with $L_1$ so does $k-L$.
This means that
the action of the defect on the NS-component of the oriented boundary 
state with $L={k\over 2}$ produces a sum with unit coefficients
of unoriented boundary states with
$L\neq k/2$. For odd $L_1$, this is already the full story, since $k/2$
does not appear in the fusion of $k/2$ with $L_1$.
Furthermore, defects with odd $L_1$ annihilate the RR-component of the boundary
state due to the $\widehat{\mathfrak{su}}(2)_k$-part of the
$S$-matrix. Hence
for $L_1$ odd
\beq\label{defLodd}
{\mathcal D}_{[L_1,M_1,S_1,\bar S_1]} \kket{[\frac{k}{2},M_2,S_2]}_B
= \frac{1}{2}
\sum_L {\mathcal N}_{L_1 \frac{k}{2}}^L \kket{[L,M_1+M_2,S_1+\bar
  S_1+S_2]}_B\,.
\eeq
On the other hand, if $L_1$ is even, the fusion of $L_1$ with $k/2$ will
again contain $k/2$, and instead of annihilating the RR-component of
the boundary state, the defect operator multiplies
it the respective Ishibashi states $\iket{[{k\over 2},{k+2\over
2},s]}_B$ by 
$(-1)^{L_1+M_1-(S_1+\bar S_1)s\over 2}$. 
(Recall $L_1$ and therefore $M_1-S_1-\bar S_1$ are even, and $s$ is odd.)
It is then clear that the defect will change the spin structure
according to $(S_2\;{\rm mod}\;2)\mapsto (S_1+\bar S_1+S_2\;{\rm
  mod}\;2)$. Whether the resulting boundary state has 
$S$-label $S_1+\bar S_1+S_2$
or $S_1+\bar S_1+S_2+2$ is determined by the overall sign of the
RR-component. Altogether, for $L_1$ even one arrives at 
\beqa\label{defLeven}
&&{\mathcal D}_{[L_1,M_1,S_1,\bar S_1]} \kket{[\frac{k}{2},M_2,S_2]}_B
= \frac{1}{2}
\sum_{L\neq k/2} 
{\mathcal N}_{L_1 \frac{k}{2}}^L \kket{[L,M_1+M_2,S_1+\bar S_1+S_2]}_B \phantom{***}\\
&&\qquad\qquad + \kket{[\frac{k}{2},M_1+M_2, S_1+\bar
  S_1+(-1)^{\scriptstyle S_1+\bar S_1}S_2-L_1-M_1-(S_1+\bar S_1)^2]}_B\,.\nonumber
\eeqa
The orientation of the oriented boundary state appearing in this
composition depends on the defect labels and $S_2$ in a rather complicated way. 
In the case that the defect preserves the spin structure of the
boundary state (that is $S_1+\bar{S}_1$ is even, and hence also $L_1+M_1$ even), the
corresponding $S$-label is given by $S_1+\bar S_1+S_2-L_1-M_1$. 
If on the other hand the defect changes the spin-structure, \ie
$S_1+\bar S_1$ is odd, the resulting $S$-label becomes
$S_1+\bar S_1-S_2-L_1-M_1-1$. 

\medskip
\noindent
{\bf A-type boundary states}
\medskip

\noindent
Let us close the discussion of the conformal field theory 
of topological defects by noting that since these defects are topological, they also act naturally
on A-type boundary states. These satisfy gluing relations
\beqn\
&&(L_n-\overline L_{-n})\kket{A}=0\,.\\
&&(G^\pm_r -i\eta \overline G^\mp_{-r})\kket{A}=0\nonumber
\eeqn
and, in the theory corresponding to $W=X^d+Z^2$, 
are given by the standard Cardy boundary
states
\beqn\label{abdstates}
\kket{[L,M,S]}_A^{NS} &=& \sqrt{2} \sum_{\stackrel{[l,m,s]}{s\;{\rm even}}}
\frac{S_{[L,M,S][l,m,s]}}{\sqrt{S_{[0,0,0][l,m,s]}}} \iket{[l,m,s]}_A
\,,\\
\kket{[L,M,S]}_A^{R} &=& \sqrt{2} \sum_{\stackrel{[l,m,s]}{s\;{\rm odd}}}
\frac{S_{[L,M,S][l,m,s]}}{\sqrt{S_{[0,0,0][l,m,s]}}} \iket{[l,m,s]}_A
\,.\nonumber
\eeqn
The discussion of the action of the topological 
defects on these boundary states is similar to the one for the B-type 
boundary states, with the result
\beq\label{adefaction}
{\mathcal D}_{[ L_1, M_1, S_1,\bar S_1]}\kket{[ L_2, M_2, S_2]}_A
=\sum_{ L}{\mathcal N}_{ L_1 L_2}^{ L} \kket{[
L, M_1+ M_2, S_1+\bar S_1+ S_2]}_A\,.
\eeq
We omit a discussion for the A-type boundary states in the theory
with the other definition of the fermion number. Let us just mention
that one can again use orbifold techniques to construct the boundary
states from the ones given above. Since none of the states \eq{abdstates}
is invariant under the respective orbifold group, they do not get
twisted sector contributions in the orbifolding construction. Instead
they can all be represented as orbits under the orbifold group 
of the boundary states in the unorbifolded theory. Hence, the action of
the defects on these boundary conditions can be easily deduced from
\eq{adefaction}.
\subsubsection{Comparison to the Landau-Ginzburg analysis}
We can now compare the results 
to the Landau-Ginzburg analysis of
Section \ref{LGapproach}. 
Matrix factorisations of type $P_I$ for $I$
consisting of consecutive integers modulo $d$ are known to correspond
to permutation boundary conditions in the tensor product of minimal
models \cite{Brunner:2005fv,Enger:2005jk}. The folding trick therefore
implies that the $P_I$ indeed correspond to the topological defects
constructed above. More precisely:
\beqn
P_{\{m,m+1,\ldots,m+l\}} &\leftrightarrow& {\mathcal D}_{[l,l+2m,0,0]}\,,\\
P^\pm_{\{m,m+1,\ldots,m+l\}} &\leftrightarrow& {\mathcal
  D}_{[l,l+2m,1\mp 1,0]}\,.\nonumber
\eeqn
The folding trick guarantees that this identification is compatible
with the topological spectra. Comparing the formula \eq{gentopcomp}
for compositions of the defects $P_I$ in 
Landau-Ginzburg models with the formula \eq{CFTdefcomp} for the composition of the
topological defects ${\mathcal D}_{[l,l+2m,S,0]}$, $S\in\{0,2\}$ in minimal models 
one indeed also finds agreement.


Using the correspondence \cite{Brunner:2003dc}
\beq
T_l \leftrightarrow \kket{[l-1,l-1,0]}_B
\eeq
between matrix factorisations of $W=X^d$ and boundary conditions in
minimal models,
one easily observes that the agreement found for
the composition of defects also holds for
the action of defects on boundary conditions (\cf equations
\eq{gentopact} and \eq{CFTdefoperation}).

This extends to the defect action on the tensor product factorisations
$\Theta_l$, in models with superpotentials $W=X^d+Z^2$.
As has been discussed in \cite{Kapustin:2003rc} 
they correspond  to the unoriented ``long orbit'' boundary states with
$L\neq{k\over 2}$ in \eq{Bboundz2}:
\beqn
\Theta_l&\leftrightarrow& \kket{[l-1,l-1,0]}_B\quad{\rm for}\;l\neq
{d\over 2}\,,\\
\Theta_{d\over 2}&\leftrightarrow& \kket{[{d\over 2}-1,{d\over
    2}-1,0]}_B+\kket{[{d\over 2}-1,{d\over 2}-1,2]}_B\,,\nonumber
\eeqn
and a comparison between \eq{gentopactz2} and \eq{deflongorbit} shows
agreement for the defect action on these. The matrix factorisations
$\Psi^\pm$ on the other hand which exist for even $d$ correspond to 
the oriented ``short orbit'' boundary
states with $L={k\over 2}$ in \eq{Bboundz2} \cite{Kapustin:2003rc}
\beq
\Psi^\pm\leftrightarrow \kket{[{d\over 2}-1,{d\over 2}-1,1\mp 1]}_B\,.
\eeq
Also for these boundary conditions the defect action derived in the
Landau-Ginzburg framework \eq{defleven}, \eq{deflodd} agrees with the
one found in the conformal field theory \eq{defLeven}, \eq{defLodd}. 

Let us close this discussion by noting that the matrix factorisations
$T_{i,j}$ are indeed tensor product matrix factorisations\footnote{We
  discuss these types of defects in the model corresponding to
  $W=X^d$. The discussion immediately carries over to the
  model associated to $W=X^d+Z^2$.}. The latter
are known to correspond to the respective tensor product boundary
states in the IR. The folding trick therefore implies the
identification of these matrix factorisations with the completely
reflective conformal B-type defects
\beq
T_{i,j}\leftrightarrow \kket{[i-1,i-1,0]}\bbra{[j-1,j-1,0]}
\eeq
in the minimal models ${\mathcal M}_k$. That this identification is
compatible with the topological spectra is clear from the folding
trick. Since these defects are not topological their composition and
action on boundary conditions is not well-defined in the CFT.
\section{Discussion}
In this paper, we have discussed B-type defects in the context of 
Landau-Ginzburg theories. Those defects between models with
superpotential $W_1$ and $W_2$ can be described by matrix
factorisations of $W_1-W_2$. We have discussed how 
two such defects can merge, and how they act on B-type boundary
conditions, which in turn have a description in terms of matrix factorisations
of the individual superpotentials $W_i$.
These two operations turn out to be quite similar, namely, they are both given
by taking the tensor product of the  matrix factorisations
describing defects and boundary conditions respectively. 
The resulting factorisations are a priori infinite 
dimensional, but can be reduced to finite dimensional ones by
splitting off infinitely many brane-anti-brane pairs. We have
described a method of how to obtain the reduced factorisations without
going through the explicit reduction procedure.

We have discussed the special defects arising from symmetries
of the bulk theories, and compared in detail the
description of B-type defects in Landau-Ginzburg models with
superpotentials $W=X^d$, $W=X^d+Z^2$, with the one of defects
in the corresponding IR CFTs.

As a next step it would now be interesting to extend the analysis
to charge-projected Landau-Ginzburg models with several superfields,
which in the IR flow to superconformal field theories with $c=9$ and 
describe the stringy regime of Calabi-Yau compactifications.
Of course, the matrix factorisations for
Landau-Ginzburg models with more chiral superfields are more
complicated, but at least for models where the superpotential 
is a Fermat polynomial the
factorisations described here can be used as building blocks.
Furthermore, the orbifold construction introduces more structure,
because it makes it necessary to consider
graded matrix factorisations \cite{Ashok:2004zb,Hori:2004ja}.

Since certain orbifolds of Landau-Ginzburg models have a geometric
interpretation as sigma model with target space $X$, the
projective variety defined by the vanishing of the superpotential, the
question about the geometric realisation of the defects and the D-branes
they act on arises. For D-branes, the connection between matrix 
factorisations and large volume geometry has been investigated in 
\cite{HHP,Aspinwall:2006ib,Aspinwall:2007cs,
Ashok:2004xq,Walcher:2004tx,Govindarajan:2005im}.
A first idea of a geometric realisation of the defects can be obtained
via the folding trick, according to which defects connecting two
sigma models with target spaces $X,Y$ 
correspond to B-type D-branes on the product 
$X\times Y$. The respective D-brane category in the topologically
twisted theory can be described by $D^\flat(Coh(X\times Y))$, 
the derived category of coherent sheaves on the product space.
 
According to our general discussion, we expect that the defects act on D-branes
and hence should provide transformations from $D^\flat (Coh(X))$ to 
$D^\flat (Coh(Y))$,
and indeed one can associate to any element $\Phi \in D^\flat (Coh(X\times Y))$ 
a Fourier-Mukai transformation with kernel $\Phi$ \footnote{A 
Landau-Ginzburg realisation
of certain Fourier-Mukai transformations , namely monodromy actions, has been
discussed in  
\cite{Jockers:2006sm}.}. 
Conversely, it has been shown
\cite{OrlovFMT,BondalOrlov} 
that any equivalence $D^\flat (Coh(X)) \to D^\flat (Coh(Y)) $ can be 
written as a Fourier-
Mukai transformation. It therefore seems plausible that defects have
a natural interpretation as Fourier-Mukai transformations at large volume. 

For some simple transformations this is indeed the case. For instance,
in Landau-Ginzburg orbifolds, there is a quantum symmetry which is
broken once one moves away from the Landau-Ginzburg point in the bulk
moduli space. We can associate a symmetry defect to this operation, which acts on the D-branes
in the Landau-Ginzburg model.
This quantum symmetry is
known to correspond to the B-brane monodromy transformation around the
Landau-Ginzburg point, which in the geometric context can be realised by a
Fourier-Mukai transformation.
We hope to come back to this issue in the future.
\section*{Acknowledgements}
We would like to thank Costas Bachas, Hans  Jockers and 
Matthias Gaberdiel for discussions.
DR would like to thank ETH Z\"urich for hospitality, where part of this work was done.
The work of IB is supported by a EURYI award and the Marie-Curie network
Forces Universe (MRTN-CT-2004-005104). DR was partially supported by 
DOE grant DE-FG02-96ER40959.
\appendix
\section{Spectra of symmetry defects}\label{sectglspectra}
Let $W(X_i)$ be a polynomial in the variables $X_1,\ldots,X_n$. Furthermore let
$\Gamma$ be a group acting linearly and unitarily on the space spanned by the
$X_i$. 
Then for any $g\in\Gamma$, the polynomial $W(X_i)-W(Y_i)$ can be
written as in 
\eq{groupfactor} leading to the matrix factorisation $D_g$ of \eq{gmf}. Here we
would like to outline the calculation of the BRST-cohomology
$\HH^*(D_g,D_1)$. For this let $Q:=D_g$ and
$R:=\CC[X_i,Y_i]/(W(X_i)-W(Y_i))$. As used in Section \ref{sectdefmf}, 
\beq
\HH^i(Q,D_1)\cong\Ext^{2+2n+i}_R(\coker(q_1),R/(X_i-Y_i))\,.
\eeq
The $\Ext$-groups can be calculated as the cohomology of the sequence
obtained by applying the functor $\Hom(\cdot,R/(X_i-Y_i))$ to the 
the $R$-free resolution of $\coker(q_1)$ given by the
matrix factorisation $Q$. But this sequence can be written as
\beq
\ldots\stackrel{\wt q_1}{\longrightarrow}\left(R/(X_i-Y_i)\right)^{2^n}
\stackrel{\wt q_0}{\longrightarrow}\left(R/(X_i-Y_i)\right)^{2^n}
\stackrel{\wt q_1}{\longrightarrow}\left(R/(X_i-Y_i)\right)^{2^n}
\longrightarrow 0\,,
\eeq
where $\wt q_a=q_a(X_j,Y_j=X_j)$. Let us assume that $g$ acts
diagonally on the $X_i$. Then, setting $Y_j=X_j$ in each of
the tensor factors $P^i$ of $D_g$ amounts to $\wt p_1^i=0$, $\wt
p_0^i=A_i(X_j,Y_j=X_j)$ in case $X_i$ is $g$-invariant, and
$\wt p_1^i=(1-g)(X_i)$, $\wt p_0^i=0$ otherwise\footnote{The latter  
can always be achieved by means of an equivalence transformation.}.  
From this it is obvious that $\ker(\wt q_a)$ is non-trivial only
if $a+|\{j|X_j\neq g(X_j)\}|$ is even, in which case the kernel is
just 
\beq
\ker(\wt q_a)\cong R/(X_i-Y_i)R
\eeq
and 
\beqn
\im(\wt q_{a+1})&=&\sum_{X_j\neq g(X_j)} (1-g)(X_j)R/(X_i-Y_i)R\\
&&\qquad\qquad +\sum_{X_j=g(X_j)}A_j(X_i,Y_i=g(X_i))R/(X_i-Y_i)R\,.\nonumber
\eeqn
Thus for $a=|\{j|X_j\neq g(X_j)\}|=:N_{\rm n-inv}$
\beq
\ker(\wt q_a)/\im(\wt q_{a+1})\cong\CC[X_j^{\rm inv}]/(\partial_j
W_{\rm inv})\,,
\eeq
where $X_j^{\rm inv}$ are the $g$-invariant variables and $W_{\rm inv}$
is obtained from $W$ by setting all non-invariant variables to zero.
Therefore we obtain
\beqn
\HH^{N_{\rm n-inv}}(D_g,D_1)&\cong& \CC[X_j^{\rm inv}]/(\partial_j
W_{\rm inv})\,,\\
\HH^{N_{\rm n-inv}+1}(D_g,D_1)&\cong& \{0\}\,.\nonumber
\eeqn
The result can be 
summarised as follows: every state in $\HH^*(D_g,D_1)$ can be written
as $p(X_i^{\rm inv})\prod_{j:X_j\neq g(X_j)}\omega_j$,
where $\omega_j$ are fermions associated to every non-$g$-invariant
variable $X_j$, and $p\in\CC[X_i^{\rm inv}]/(\partial_i W_{\rm inv})$
is a polynomial in the $g$-invariant variables $X_j^{\rm
inv}=g(X_j^{\rm inv})$. $W_{\rm inv}$ is obtained from $W$ by setting
all non-invariant variables to zero. 
This is in agreement with the $g$-twisted bulk Hilbert spaces obtained
in \cite{Intriligator:1990ua,Vafa:1989xc}.

\section{Explicit equivalence for $D_1\otimes T_1$}\label{explicitequiv}
Here we would like to show explicitly that the infinite dimensional
matrix factorisation
\beq
D_1\otimes T_1(Y):\quad r_1=\left(\begin{array}{cc} X-Y & -Y\\ Y^{d-1} & {X^d-Y^d\over
X-Y}\end{array}\right)\,,\quad
r_0=\left(\begin{array}{cc} {X^d-Y^d\over X-Y} & Y\\ -Y^{d-1} & X-Y\end{array}\right)
\eeq
of $X^d$ over $\CC[X]$, which is obtained as the tensor product
of the matrix factorisations
\beq
D_1:\quad p_1=(X-Y),\; p_0={X^d-Y^d\over X-Y}
\eeq
of $X^d-Y^d$ and 
\beq
T_1(Y):\quad q_1=Y,\; q_0=Y^{d-1}
\eeq
of $Y^d$ is indeed equivalent to $T_1(X)$. Using the trick discussed
in Section \ref{sectdefmf} one easily arrives at this conclusion,
because $D_1\otimes T_1(Y)$ has to be equivalent to the matrix
factorisation obtained from the $R=\CC[X]/(X^d)$-free resolution of 
the module $M:=\coker(X-Y,Y)\cong\coker(X,Y)\cong R/XR$ by
chopping off an even number of terms. But obviously $M$ has an $R$-free
resolution given by $T_1(X)$. 

To construct the equivalence explicitly note first that by means of 
\beqn
&&u_0=\left(\begin{array}{cc} 1 & 0 \\ {1\over X}\left({X^d-Y^d\over
X-Y}-Y^{d-1}\right) & 1\end{array}\right)\,,\quad v_0=u_0^{-1}\,,\\
&&u_1=\left(\begin{array}{cc} 1 & 0 \\ -1 &
-1\end{array}\right)\,,\quad v_1=u_1^{-1}
\eeqn
$(r_1,r_0)$ is equivalent to
\beq
r\p_1=\left(\begin{array}{cc} X& Y\\ 0 & -X^{d-1}\end{array}\right)\,,\quad
r\p_0=\left(\begin{array}{cc} X^{d-1} & Y\\ 0 &
-X\end{array}\right)\,.
\eeq
Regarding $\CC[X,Y]$ as the infinite dimensional free $\CC[X]$-module
$\CC[X,Y]\cong\CC[X]+Y\CC[X]+Y^2\CC[X]+\ldots$, $Y$ can be represented by the
infinite dimensional matrix
\beq
Y=\left(\begin{array}{ccc} 0 & &  \\ 1 & \ddots &  \\ & \ddots & \ddots 
 \end{array}\right)\,.
\eeq
Using this representation $r\p_1$ takes the form
\beq
r\p_1=\left(\begin{array}{ccc|ccc} 
X & & &        0 & &\\
  & \ddots & & 1 & \ddots & \\
&&\ddots     &   & \ddots &\ddots \\
\hline
&&& -X^{d-1}&&\\
&&& &\ddots &\\
&&&&&\ddots
\end{array}\right)
\eeq
Now one easily finds the following chain of elementary
row and column transformations of $r\p_1$:
\beqn
r\p_1&\mapsto&
\left(\begin{array}{ccc|ccc} 
X & & &        0 & &\\
  & 1 & & X & \ddots & \\
&&\ddots     &   & \ddots &\ddots \\
\hline
0&-X^{d-1}&& &&\\
&\ddots&\ddots& & &\\
&&\ddots&&&
\end{array}\right)
\mapsto
\left(\begin{array}{ccc|ccc} 
X & & &        & &\\
  & 1 & & &  & \\
&&\ddots     &   &  & \\
\hline
0&-X^{d-1}&& X^d &&\\
&\ddots&\ddots& &\ddots &\\
&&\ddots&&&\ddots
\end{array}\right)\nonumber\\
&\mapsto&
\left(\begin{array}{ccc|ccc} 
X & & &         & &\\
  & 1 & & &  & \\
&&\ddots     &   &  & \\
\hline
&&& X^d &&\\
&&& &\ddots &\\
&&&&&\ddots
\end{array}\right)
\eeqn
The opposite transformations lead to 
\beq
r\p_0\mapsto
\left(\begin{array}{ccc|ccc} 
X^{d-1} & & &         & &\\
  & X^d & & &  & \\
&&\ddots     &   &  & \\
\hline
&&& 1 &&\\
&&& &\ddots &\\
&&&&&\ddots
\end{array}\right)\,,
\eeq
and hence we have obtained an explicit equivalence of the infinite
dimensional matrix factorisation $D_1\otimes T_1(Y)$ to the sum of
the matrix factorisation $T_1(X)$ with infinitely many trivial matrix
factorisations $(1,X^d)$. 
%
%
\bibliographystyle{lgdef}
\bibliography{lgdef}
\end{document}